\documentclass[%
 reprint,
superscriptaddress,
nofootinbib,
 amsmath,amssymb,
 aps,
]{revtex4-1}
\usepackage{graphics,epsfig,psfrag}
\usepackage{color}
\usepackage{graphicx}% Include figure files
\usepackage{dcolumn}% Align table columns on decimal point
\usepackage{bm}% bold math
\begin{document}

\preprint{APS/123-QED}

\title{
Light propagation in a plasma on Kerr spacetime: Separation of the 
Hamilton-Jacobi equation and calculation of the shadow}

\author{Volker Perlick}
\email{perlick@zarm.uni-bremen.de}
\affiliation{ZARM, University of Bremen, 28359 Bremen,
Germany}

\author{Oleg Yu. Tsupko}
\email{tsupko@iki.rssi.ru}
\affiliation{Space Research Institute of Russian Academy of Sciences, Profsoyuznaya 84/32, Moscow 117997, Russia}
\affiliation{National Research Nuclear University MEPhI (Moscow Engineering Physics Institute), \\Kashirskoe Shosse 31, Moscow 115409, Russia}%

\date{\today}% It is always \today, today,

\begin{abstract}
We consider light propagation in a non-magnetized pressureless plasma 
around a Kerr black hole. We find the necessary and sufficient condition
the plasma electron density has to satisfy to guarantee that the 
Hamilton-Jacobi equation for the light rays is separable, i.e., that
a generalized Carter constant exists. For all cases where this condition
is satisfied we determine the photon region, i.e., the region in the 
spacetime where spherical light rays exist. A spherical light ray is a 
light ray that stays on a sphere $r = \mathrm{constant}$ (in Boyer-Lindquist
coordinates). Based on these results, we calculate the shadow of a Kerr 
black hole under the influence of a plasma that satisfies the separability 
condition. More precisely, we derive an analytical formula for the 
boundary curve of the shadow on the sky of an observer that is located
anywhere in the domain of outer communication. Several
examples are worked out.

\begin{description}
\item[PACS numbers] 04.20.-q -- 98.35.Jk -- 98.62.Mw -- 98.62.Sb
%\pacs{#04.20.-q}

% - 95.30.Sf - 98.62.Sb - 94.20.ws}

\end{description}
\pacs{?????? - ??????}
\end{abstract}

\pacs{?????? - ??????}% PACS, the Physics and Astronomy
\maketitle

%\tableofcontents

\section{Introduction}\label{sec:intro}

For most applications of general relativity light rays can be 
considered as lightlike geodesics of the spacetime metric, i.e.,
the influence of a medium on the light rays can be neglected. 
However, in the radio frequency range this is not always true. A well-known example is the influence of the Solar corona on
the travel time and on the deflection angle of radio signals that
come close to the Sun. This influence is routinely observed 
since the 1960s. 
In this case one may assume that the medium is a non-magnetized 
pressureless plasma and for the gravitational field the 
linearized theory is sufficient. The relevant equations have
been determined by Muhleman et al. 
\cite{MuhlemanJohnston1966,MuhlemanEkersFomalont1970}. 
Gravitational lensing in this approximation was firstly 
discussed by Bliokh and Minakov \cite{BliokhMinakov}. In 
these works both the gravitational and the plasma deflection 
are assumed to be small and they are calculated separately 
from each other.

There is good reason to assume that also black holes and other
compact objects are surrounded by a plasma and it is an interesting
question to investigate if the plasma could have an observable effect
on radio signals that come close to such a compact object. In such cases the
linearized theory is not sufficient. For a spherically symmetric
uncharged black hole, one has to consider the Schwarzschild
metric and for a rotating uncharged black hole one has to consider the
Kerr metric. Some results on light propagation in a non-magnetized pressureless
plasma on the Schwarzschild or Kerr spacetime are known. The influence of a 
spherically symmetric and time-independent plasma density on the light 
deflection in the Schwarzschild spacetime was 
calculated and discussed by Perlick \cite{Perlick2000}. 
In this study neither the gravitational nor the plasma deflection
is assumed to be small. Later, the influence of plasma effects on 
gravitational lensing was investigated,
with different methods, by Bisnovatyi-Kogan and Tsupko 
\cite{BKTs2009, BKTs2010, TsBK2013}. In the Kerr spacetime, light deflection in the
equatorial plane was calculated by Perlick \cite{Perlick2000} for
the case of a rotationally symmetric and time-independent plasma
density. Within the approximation of small deflection, lensing of light rays off
the equatorial plane was considered by 
Morozova et al. \cite{MorozovaAhmedovTursunov2013}
who assumed a slowly rotating Kerr black hole. The influence of a plasma
on the multiple imaging properties in the strong-bending regime was
investigated by Bisnovatyi-Kogan and Tsupko \cite{TsBK2013} for the 
Schwarzschild spacetime and recently extended to Kerr spacetime by 
Liu, Ding and Jing \cite{LiuDingJing2016}. For some
possible astrophysical observations of plasma effects near compact 
objects we refer to 
Rogers \cite{Rogers2015,Rogers2017a,Rogers2017b} and for plasma 
effects in strong lens systems to Er and Mao \cite{ErMao2014}. For a
review see \cite{BKTs2015}.

Here we want to concentrate on the influence of a plasma on the shadow
of a black hole. Roughly speaking, the shadow is the black disk an observer
sees in the sky if a black hole is viewed against a backdrop of light
sources that are distributed around the black hole but not between the
observer and the black hole. For constructing the shadow 
we have to consider all past-oriented light rays that issue from a
chosen observer position. Each of these light rays corresponds to a 
point on the observer's sky. We assign darkness to a point if the corresponding
light ray goes to the horizon and brightness otherwise.
The idea is that there are light sources distributed in
the spacetime, but not between the observer and the
black hole and not inside a possible white-hole extension
of the spacetime from where future-oriented light rays
could be sent across the horizon towards the observer. Without a plasma, the shadow of a 
Schwarzschild black hole was calculated by Synge \cite{Synge1966} and the shape of the
shadow of a Kerr black hole, for an observer at infinity, was calculated
by Bardeen \cite{Bardeen1973}. 
In an earlier paper \cite{PerlickTsupkoBK2015} we have generalized
Synge's formula to the case of a spherically symmetric and static 
plasma distribution on a spherically symmetric and static spacetime. 
As particular examples, we have worked out the
results for the case that the underlying spacetime geometry is 
(a) that of a Schwarzschild black 
hole and (b) that of an Ellis wormhole. In the 
present paper we want to derive a similar result for a plasma around a
Kerr black hole. More precisely, it is our goal to derive an analytical
formula for the boundary curve of the shadow on the sky of an observer at
an arbitrary position outside of the horizon of the black hole (i.e.,
in contrast to Bardeen we do not consider an observer at infinity). 
Without the influence of a plasma, such a formula has been derived
by Grenzebach et al. 
\cite{GrenzebachPerlickLaemmerzahl2014,GrenzebachPerlickLaemmerzahl2015}
for the case of a black hole of the Pleba{\'n}ski-Demia{\'n}ski
class (which includes the Kerr black hole as a special case). Their
method made use of the fact that the equation of motion for 
vacuum light rays admits a fourth constant of motion, the Carter
constant, in addition to the ones that follow from the symmetry. Therefore,
an important first step in our analysis will be to find out under
what conditions a Carter constant exists for light propagation in a
plasma on a Kerr spacetime. Of course, this result is of interest not
only for the calculation of the shadow but also for other problems.

The perspectives of actually observing the influence of a plasma on the
shadow, e.g. for Sgr A$^*$ or for M87, have been discussed in our earlier
paper \cite{PerlickTsupkoBK2015}.
 
The paper is organized as follows. In Sec.~\ref{sec:Hamilton} we review
the Hamilton formalism for light rays in a non-magnetized, pressureless plasma
on a general-relativistic spacetime and we specialize the relevant equations 
to the Kerr metric. In Sec.~\ref{sec:sepcon} we derive the necessary and
sufficient condition on the plasma electron density that guarantees 
separability of the Hamilton-Jacobi equation for light rays, i.e., that
guarantees the existence of a Carter constant\footnote{Atamurotov et al. \cite{Ahmedov2015} discuss the shadow of a 
Kerr black hole in a plasma whose density depends on $r$ only. They 
assume that in this case a Carter constant exists. In Section \ref{sec:sepcon}
we will demonstrate that this is actually not the case: For a Kerr black
hole with $a \neq 0$ and a non-zero plasma density that depends on $r$ only,
the Hamilton-Jacobi equation for the light rays is \emph{not} separable, i.e.,
there is no Carter constant. Therefore, the pictures
of the Kerr shadow in a plasma 
produced in Ref. \cite{Ahmedov2015} are, unfortunately, based
on incorrect equations.}. 
In Sec.~\ref{sec:phreg}
we determine the photon region for a plasma density around a Kerr black
hole that satisfies the separability condition. The photon region is the
spacetime region filled with spherical light rays, i.e., with light rays 
that stay on a sphere $r = \mathrm{constant}$ in Boyer-Lindquist coordinates.
Knowledge on the location of the spherical geodesics and on the constants
of motion associated with them is of crucial relevance for deriving the
boundary curve of the shadow on the sky of an observer  anywhere in the 
domain of outer communication of the black hole. This will be done, in
close analogy to the procedure in Grenzebach et 
al.~\cite{GrenzebachPerlickLaemmerzahl2014,GrenzebachPerlickLaemmerzahl2015},
in Sec.~\ref{sec:shadow}. In the rest of the paper we determine the 
photon regions and the shadow for some special cases where the separability
condition is satisfied.

Our conventions are as follows. We use the summation convention for
greek indices that take the values 0,1,2,3.  Our choice of signature is 
$(-,+,+,+)$. We raise and lower greek indices with the spacetime
metric. $G$ is the gravitational constant and $c$
is the vacuum speed of light. We use units such that $\hbar = 1$, i.e,
energies have the same unit as frequencies and momentum vectors have the
same unit as wave vectors.

%-----------------------------------------------------------------------------
\section{Hamilton formalism for light rays in a plasma on Kerr spacetime}\label{sec:Hamilton}

Light propagation in a non-magnetized pressureless plasma can be 
characterized by the Hamiltonian
\begin{equation}\label{eq:Hg}
H (x, p) = \dfrac{1}{2} \big( g^{\mu \nu} (x) p_{\mu} p_{\nu} 
+ \omega _p (x) ^2 \big)
\, .
\end{equation}
Here $g^{\mu \nu}$ are the contravariant components of the spacetime
metric tensor and $\omega _p$ is the plasma electron frequency which equals, up
to a scalar factor, the electron density, 
\begin{equation}\label{eq:omega_p}
\omega_p(x) = \frac{4 \pi e^2}{m_e} N_e(x)
\end{equation}
where $e$ and $m_e$ are the electron charge and mass, respectively, 
and $N_e$ is the electron number density.
$x = (x^0,x^1,x^2,x^3)$ are 
the spacetime coordinates and $p=(p_0,p_1,p_2,p_3)$ are the 
canonical momentum coordinates. The light rays are the solutions to 
Hamilton's equations
\begin{equation}\label{eq:Ham}
\frac{dx^{\mu}}{ds} = \frac{\partial H}{\partial p_{\mu}} \, , \quad
\frac{dp_{\mu}}{ds} = - \, \frac{\partial H}{\partial x^{\mu}} \, , \quad
H(x,p) = 0 \, ,
\end{equation}
where $s$ is a curve parameter which has no direct physical or geometrical
meaning. A rigorous derivation from Maxwell's equations of this Hamiltonian
approach was given, even for the more general case of a magnetized
plasma, by Breuer and Ehlers \cite{BreuerEhlers1980,BreuerEhlers1981}.
For the much simpler case of a non-magnetized plasma, a similar
derivation can be found in \cite{Perlick2000}. 

In regions where $\omega _p \neq 0$ we may use the Hamiltonian
\begin{equation}\label{eq:confH}
\tilde{H} (x,p) = \dfrac{1}{2}
\Big( \dfrac{c^2}{\omega _p (x) ^2} \, g^{\mu \nu} (x) p_{\mu} p_{\nu}
+ c^2 \Big) \, = 
\end{equation}
\[
= \, \dfrac{c^2}{\omega _p (x) ^2} \, H( x , p )
\]
because multiplying a Hamiltonian with a nowhere vanishing function does not
affect the solution curves to (\ref{eq:Ham}), except for a reparametrization 
$s \mapsto \tilde{s}$. Using the Hamil\-tonian $\tilde{H}$ demonstrates 
that the rays are \emph{timelike} geodesics of the conformally rescaled 
metric $\tilde{g}{}_{\mu \nu} dx^{\mu} dx^{\nu} = 
c^{-2} \omega _p ^2 \, g_{\mu \nu} dx^{\mu}dx^{\nu}$. The new 
curve parameter $\tilde{s}$ is proper time with respect to this metric. 
Although such a conformal rescaling is often convenient, in the following we 
prefer to work with the Hamiltonian $H$, rather than with the
Hamiltonian $\tilde{H}$, because then the equations can be immediately
applied to all regions, independently of whether $\omega _p \neq 0$
or $\omega _p = 0$.

A plasma is a dispersive medium, i.e., light propagation depends on the 
frequency. 
To assign a frequency to a light ray, we have to choose an 
observer with four-velocity $U^{\mu} (x)$, normalized to 
$g_{\mu \nu} (x) U^{\mu} (x) U^{\nu} (x) = - c^2$. Decomposing the
momentum into a component parallel to $U^{\mu} (x)$ and a component
orthogonal to $U^{\mu} (x)$,
\begin{equation}\label{eq:pmudec}
p^{\mu} = - \, \frac{1}{c} \, \omega (x) \, U^{\mu} (x) + k^{\mu} (x) \, ,
\end{equation}
defines the frequency
\begin{equation}\label{eq:freq}
\omega (x) =  \frac{1}{c} \, p_{\mu} U^{\mu} (x) 
\end{equation}
and the spatial wave vector 
\begin{equation}\label{eq:kmu}
k^{\mu} (x) = p^{\mu} + \frac{1}{c^2} \, p_{\nu} U^{\nu} (x) \, U^{\mu} (x)  \, .
\end{equation}
Note that $\omega (x)$ is positive if $U^{\mu} (x)$ is future-pointing and 
$p^{\mu}$ is past-pointing. We have chosen this somewhat unusual convention
because later, when we calculate the shadow, we will consider light rays
issuing from the observer position into the past and we want to assign a
positive frequency to those rays.

With the decomposition (\ref{eq:pmudec}), the condition $H=0$ may be solved
for the frequency which gives the dispersion relation in the familiar
form,
\begin{equation}\label{eq:freqcon}
\omega (x)^2  = k_{\mu} (x) k^{\mu} (x) + \omega _p (x) ^2 \, .
\end{equation}
As $k^{\mu} (x)$ is spacelike, this equation implies
\begin{equation}\label{eq:omegacon}
\omega (x) ^2 \ge \omega _p (x) ^2 \, ,
\end{equation}
i.e., light propagation is possible only with a frequency that is bigger 
than the plasma frequency. If we introduce in the usual way the 
phase velocity 
\begin{equation}\label{eq:vpdef}
v_p \big( x , \omega (x) \big) = \sqrt{\frac{\omega (x) ^2 }{k_{\mu} (x) k^{\mu} (x)}}
\end{equation}
and the index of refraction 
\begin{equation}\label{eq:ndef}
n \big( x , \omega (x) \big)  = 
\frac{c}{v_p \big( x , \omega (x) \big)} \, ,
\end{equation}
we see that the index of refraction, as a function of $x$ and $\omega (x)$, is of the
same form for all $U^{\mu} (x)$,
\begin{equation}\label{eq:n}
n \big( x, \omega (x) \big) ^2 = 1 - \frac{\omega _p (x)^2}{\omega (x) ^2}
\, .
\end{equation}
The remarkable fact is that the index of refraction depends only
on the function $\omega _p$, i.e., on the plasma electron density, but not on
the four-velocity field of the plasma.
This demonstrates that a non-magnetized pressureless plasma is an example for the type
of dispersive medium that was discussed in detail in the text-book by 
Synge \cite{Synge1960}, Chapter XI. Condition (\ref{eq:omegacon}) guarantees 
that the index of refraction is real (and non-negative) for all allowed
frequencies.

We will now specialize to the Kerr metric which is given, in Boyer-Lindquist 
coordinates $x=(t, r, \vartheta , \varphi )$, by
\begin{gather}
g _{\mu \nu} dx^{\mu} dx^{\nu}  =  
 -c^2  \left(1-\frac{2mr}{\rho^2}\right) dt^2 + 
\frac{\rho^2}{\Delta} dr^2 + \rho^2 d\vartheta^2 
\nonumber
\\
\label{eq:g}
+ \, \mathrm{sin}^2\vartheta\left(r^2+a^2+
\frac{2mra^2\mathrm{sin}^2\vartheta}{\rho^2}\right)
d\varphi^2
\nonumber
\\
- \frac{4mra \mathrm{sin} ^2\vartheta}{\rho^2} \, c \, dt \, d\varphi
\end{gather}
where
\begin{equation}\label{eq:Deltarho}
\Delta = r^2 + a^2 - 2mr \, , \quad
\rho ^2 = r^2 + a^2 \mathrm{cos} ^2 \vartheta \, .
\end{equation}
Here $m$ is the mass parameter and $a$ is the spin parameter,
\begin{equation}\label{eq:mM}
m = \dfrac{GM}{c^2} \, , \quad a = \frac{J}{Mc} \, ,
\end{equation}
where $M$ is the mass and $J$ is the spin.
Throughout this paper, we assume that $a^2 \le m^2$, i.e., we consider a black hole
rather than a naked singularity. We restrict our consideration to the domain of
outer communication, i.e., to the domain outside of the outer horizon, $r >
m + \sqrt{m^2-a^2}$.

With the metric coefficients specified to the Kerr metric, the 
Hamiltonian (\ref{eq:Hg}) reads
\begin{equation}\label{eq:H2}
H = \dfrac{1}{2 \rho^2} \Bigg( - \frac{1}{\Delta}
\Big( a \, p_{\varphi} + (r^2+a^2) \frac{p_t}{c} \Big)^2
\end{equation}
\[
+ \Big(  \frac{p_{\varphi}}{\mathrm{sin} \, \vartheta}  
+ a \, \mathrm{sin} \, \vartheta \, \frac{p_t}{c} \Big) ^2
+ p_{\vartheta}^2 + \Delta \, p_r^2 + \rho ^2 \omega _p ^2 \Bigg)
\, .
\]
We assume that $\omega _p$ is a function only of $r$ and $\vartheta$. Then
$\partial H / \partial t =0$ and $\partial H / \partial \varphi =0$, i.e.
$p_t$ and $p_{\varphi}$ are constants of motion. $p_{\varphi}$ is the 
$z$ component of the angular momentum. If we write
\begin{equation}\label{eq:omega0}
p_t = c \, \omega _0 \, ,
\end{equation}
the physical meaning of the constant of motion $\omega _0$ becomes clear
if we specify the frequency (\ref{eq:freq}) to the case of an observer
on a $t$ line, $U^{\mu} (x) = c \, \delta ^{\mu}_t (-g_{tt} (x))^{-1/2}$,
\begin{equation}\label{eq:omega0t}
\omega (x) = \frac{p_t}{\sqrt{-g_{tt}(x)}} = 
\frac{\omega _0}{\sqrt{1- \dfrac{2 \, m \, r}{\rho^2 }}}
\, ,
\end{equation}
which is possible everywhere outside the ergoregion where $2 \, m \, r < \rho ^2$. 
We see that for a light ray that reaches infinity $\omega _0$ is the frequency 
measured by an observer on a $t$ line at infinity. 
In vacuo (i.e., if $\omega _p  =0$), the path of a light ray is independent of the
frequency; correspondingly, there is no restriction on $\omega _0$.
In a plasma, however, light propagation does depend on the 
frequency. If $\omega _0$ becomes too small in comparison to the plasma frequency, 
it is even impossible for a light ray with frequency $\omega _0$ to propagate 
in the plasma. 

To derive the precise form of the condition on $\omega _0$, we read from 
(\ref{eq:H2}) that the equation $H=0$ can hold at a point with coordinates
$(r, \vartheta )$ on the domain of outer communication only if
\begin{equation}\label{eq:Fpphi}
F ( p_{\varphi}) := \Big( a \, p_{\varphi} +(r^2+a^2) \omega _0 \Big)^2
\end{equation}
\[
- \Delta \Big(  \frac{p_{\varphi}}{\mathrm{sin} \, \vartheta}  
+ a \, \Delta \, \mathrm{sin} \, \vartheta \, \omega _0 \Big) ^2
- \rho ^2 \Delta \, \omega _p ^2  \ge 0 \, .
\]
Here we have used that $\Delta > 0$ on the domain of outer communication.
The function $F(p_{\varphi})$ has an extremum at
\begin{equation}\label{eq:pphi}
p_{\varphi ,e} = 
\frac{- \, 2 \, m \, r \, a \, \mathrm{sin} ^2 \vartheta \, \omega _0}{2 \, m \, r - \rho ^2}
\, .
\end{equation}
Inside the ergoregion, where $2 \, m \, r > \rho ^2$, this extremum is a minimum
and $F ( p _{\varphi} )$ takes arbitrarily large positive values, so the inequality
(\ref{eq:Fpphi}) can be satisfied for any $\omega _0$ by choosing $p_{\varphi}$
appropriately. Outside the ergoregion, where $2 \, m \, r < \rho ^2$, the extremum
is a maximum. The inequality (\ref{eq:Fpphi}) can be satisfied only if 
$F( p_{\varphi , e}) \ge 0$ which is equivalent to 
\begin{equation}\label{eq:limit2}
\omega _0 ^2 \,  \ge \, 
\Big( 1 - \dfrac{2mr}{\rho ^2} \Big) \, \omega _p (r, \vartheta ) ^2 \, .
\end{equation}
As inside the ergoregion the inequality (\ref{eq:limit2}) is true for any real 
$\omega _0$, this inequality is the necessary and sufficient condition for the 
existence of a light ray with constant of motion $\omega _0$ anywhere on the domain
of outer communication. 
If the plasma frequency is bounded on the domain outside of the 
ergoregion, $\omega _p (r , \vartheta ) \le \omega _c = \mathrm{constant}$, light
rays with $\omega _0 ^2 \ge \omega _c ^2$ can travel through any point 
of the domain of outer communication.

If we have chosen a frequency $\omega _0$ that is allowed at a point with 
coordinates $r$ and $\vartheta$, the allowed values of 
$p_{\varphi}$, $p_{\vartheta}$ and $p_{r}$
are determined by the condition $H=0$.

%-----------------------------------------------------------------------------
\section{Separability of the Hamilton-Jacobi equation for
light rays in a plasma on Kerr spacetime }\label{sec:sepcon}

If $\omega _p$ depends only on $r$ and $\vartheta$, we have three 
constants of motion for light rays, $H=0$, $p_t = c \, \omega _0$ and 
$p_{\varphi}$. We will now investigate for which special form of the
function $\omega _p (r, \vartheta )$ the Hamilton-Jacobi equation 
can be separated; in this case the separation constant will give 
us a fourth constant of motion such that the equation for light rays
becomes completely integrable. 

By (\ref{eq:H2}), the Hamilton-Jacobi equation 
\begin{equation}\label{eq:HamJac}
0 = H \Big( x , \frac{\partial S}{\partial x} \Big) 
\end{equation}
reads
\begin{gather}\label{eq:HJ}
0 =  \, - \, \frac{1}{\Delta}
\Big( a \, \frac{\partial S}{\partial \varphi}
+ (r^2+a^2) \frac{1}{c} \, \frac{\partial S}{\partial t}
 \Big)^2
\\
\nonumber
+ \Big( \frac{1}{\mathrm{sin} \, \vartheta} \, \frac{\partial S}{\partial \varphi} 
+\frac{a}{c} \, \mathrm{sin} \, \vartheta \, \frac{\partial S}{\partial t}
\Big) ^2
\\
\nonumber
+ \Big( \frac{\partial S}{\partial \vartheta} \Big)^2 
+ \Delta \, \Big( \frac{\partial S}{\partial r} \Big) ^2 
+ \rho ^2 \omega _p ^2 
\, .
\end{gather}
With the separation ansatz 
\begin{equation}\label{eq:S}
S(t, \varphi , r, \vartheta ) = 
\, c \, \omega _0  \, t + p_{\varphi} \varphi+
S_r (r) + S_{\vartheta}(\vartheta ) 
\end{equation}
the Hamilton-Jacobi equation takes the following form:
\begin{gather}
\nonumber
0 = - \dfrac{1}{\Delta} \Big( a \, p_{\varphi} + (r^2+a^2) \omega _0  \Big) ^2 
+ \Big( \dfrac{p_{\varphi}}{\mathrm{sin} \, \vartheta} + 
a \, \mathrm{sin} \, \vartheta \, \omega _0 \Big)^2
\\
\label{eq:HJ5}
+ S_{\vartheta}' ( \vartheta )^2 + \Delta S_r'(r)^2 
+ \omega _p ^2 \big( r^2 + a^2 \mathrm{cos} ^2 \vartheta \big) \, .
\end{gather}
Separability of the Hamilton-Jacobi equation requires
that \protect{(\ref{eq:HJ5})} can be rearranged in a way that the left-hand side
is independent of $r$ and the right-hand side is independent of 
$\vartheta$. We read from (\ref{eq:HJ5}) that this is possible if and
only if the plasma frequency is of the form
\begin{equation}\label{eq:sepcon}
\omega _p (r , \vartheta ) ^2 = \dfrac{f_r(r)+f_{\vartheta} ( \vartheta )}{ 
r^2 + a^2 \mathrm{cos} ^2 \vartheta}
\end{equation}
with some functions $f_r(r)$ and $f_{\vartheta} ( \vartheta )$.
Then the Hamilton-Jacobi equation reads 

\begin{gather}
S_{\vartheta}' ( \vartheta )^2
+\Big( \dfrac{p_{\varphi}}{\mathrm{sin} \, \vartheta} + 
a \, \mathrm{sin} \, \vartheta \, \omega _0 \Big)^2
+f_{\vartheta}( \vartheta )  
\nonumber
\\
\label{eq:K}
=- \Delta S_r'(r)^2
+\dfrac{1}{\Delta} \Big( a \, p_{\varphi} + (r^2+a^2) \omega _0  \Big) ^2
- f_r (r) 
\, =: \, K \, .
\end{gather}
As the first expression  is independent of $r$ whereas the second expression 
is independent of $\vartheta$, the quantity $K$ depends neither
on $r$ nor on $\vartheta$, so it is a constant. $K$ is the generalized
Carter constant.

\vspace{0.2cm}

With $S_{\vartheta} ' ( \vartheta )
= p_{\vartheta}$ and $S'_r ( r) = p_r$ we have found that
\begin{equation}\label{eq:ptheta}
p_{ \vartheta}^2 = K 
-\Big( \dfrac{p_{\varphi}}{\mathrm{sin} \, \vartheta} +
a \, \mathrm{sin} \, \vartheta \, \omega _0 \Big)^2
- f_{\vartheta} ( \vartheta ) 
\, ,
\end{equation}
\begin{equation}\label{eq:pr}
\Delta p_r^2 = - K +
\dfrac{1}{\Delta} \Big( (r^2+a^2) \omega _0 + ap_{\varphi} \Big) ^2
- f_r (r) 
\, .
\end{equation}
With these expressions for $p_{\vartheta}$ and $p_r$ inserted into
Hamilton's equations 
\begin{equation}\label{eq:Ham1}
\dot{x}{}^{\mu} = \dfrac{ \partial H}{\partial p_{\mu}}
\end{equation}
for $x^{\mu} =\vartheta$ and $x^{\mu} = r$, respectively,
we find
\begin{equation}\label{eq:dottheta}
\rho ^4 \dot{ \vartheta}{}^2 = K 
-\Big( \dfrac{p_{\varphi}}{\mathrm{sin} \, \vartheta} +
a \mathrm{sin} \, \vartheta \, \omega _0 \Big)^2
- f_{\vartheta} ( \vartheta ) 
\, ,  
\end{equation}
\begin{equation}\label{eq:dotr}
\rho ^4 \dot{r}{}^2= - K \Delta  +
 \Big( (r^2+a^2) \omega _0 + a p_{\varphi} \Big) ^2
- f_r (r) \, \Delta =:R(r)
\, . 
\end{equation}
The other two components, $x^{\mu} = \varphi$ and $x^{\mu}=t$,
of (\ref{eq:Ham1}) yield
\begin{equation}\label{eq:dotphi}
\rho ^2 \dot{\varphi}  =
\dfrac{- 2mra \, \mathrm{sin} ^2 \vartheta  \, \omega _0
+ \big( \rho^2 - 2mr \big) p_{\varphi}}{
\Delta \, \mathrm{sin}^2 \vartheta} \, , 
\end{equation}
\begin{equation}\label{eq:dott}
\rho ^2 \,  \dot{t} =
\dfrac{ - \Big(  (r^2+a^2) \rho ^2 + 2mra^2 \mathrm{sin}^2 \vartheta \Big) 
\omega _0 - 2mra p_{\varphi}}{c \, \Delta }
 \, .
\end{equation}
Formulas (\ref{eq:dottheta}), (\ref{eq:dotr})
(\ref{eq:dotphi}) and (\ref{eq:dott}) give us the 
equations of motion for the rays in first-order form. To sum up, we have demonstrated that the condition (\ref{eq:sepcon})
is necessary and sufficient for the existence of a
generalized Carter constant, and that then complete
integrability of the equations of motion is guaranteed. In the next sections we will calculate the photon regions
and the shadow. As these calculations will be based on
the existence of the Carter constant, we will have to
restrict to the case that condition (\ref{eq:sepcon}) is satisfied. For
other plasma densities our mathematical methods will
not work.

In terms of the index of refraction 
(\ref{eq:n}), the separability condition (\ref{eq:sepcon}) reads
\begin{equation}\label{eq:sepconn}
n \big( x , \omega (x) \big) ^2 = 
1 - \frac{f_r (r) + f_{\vartheta} (\vartheta )}{\omega (x)^2 \rho ^2} \, .
\end{equation}

In regions where $\omega _p \neq 0$ the rays are timelike 
curves. The curve parameter is proper time with respect to
the conformally rescaled metric $\omega _p ^{-2} g_{\mu \nu}
dx^{\mu}dx^{\nu}$. Note that a homogeneous plasma, $\omega _p
= \omega _c = \mathrm{constant}$, satisfies the separability condition
(\ref{eq:sepcon}) with 
\begin{equation}\label{eq:hom}
f_r(r) = \omega _c ^2 r^2 \, , \quad f_{\vartheta} 
(\vartheta ) = \omega _c ^2 a^2 \mathrm{cos}^2 \vartheta \, .
\end{equation}
In this case, 
the rays are timelike geodesics of the Kerr metric and the 
curve parameter is an affine parameter, i.e., it is affinely 
related to proper time of the Kerr metric. 

%We can change to
%the proper time parametrization by the transformation
%\begin{equation}\label{eq:tau}
%( \, \cdot \, ) ^{\cdot} \, \mapsto \, 
%\dfrac{\Omega}{c} ( \, \cdot \, ) ^{\cdot}
%\end{equation}
%which implies 
%\begin{equation}\label{eq:taup}
%p_{t} \, \mapsto \, \dfrac{\omega _c}{c} \, p_t 
%\, , \quad
%p_{\varphi} \, \mapsto \, \dfrac{\omega _c}{c} \, p_{\varphi}
%\, , \quad
%K \, \mapsto \, \dfrac{\omega _c ^2}{c^2} \, K
%\, .
%\end{equation}

%---------------------------------------------------------------------
\section{Photon regions in a plasma on the Kerr spacetime}\label{sec:phreg}

We consider a plasma distribution of the form of 
eq. (\ref{eq:sepcon}) so that the $\vartheta$ and the 
$r$ components of the ray equation can be written in 
separated form, see  (\ref{eq:dottheta}) and 
(\ref{eq:dotr}). 

The photon region is the region in spacetime filled
with spherical light rays, i.e., with solutions to
the ray equation that stay on a sphere $r= \mathrm{constant}$.
Unstable spherical light rays can serve as limit
curves for light rays that approach them in a 
spiral motion. In the next section we will see that, for
this reason, they are of crucial importance for the 
construction of the shadow.

Spherical light rays satify $\dot{r}=0$ and $\ddot{r}=0$, i.e.,
in the notation of (\ref{eq:dotr})
\begin{equation}\label{eq:R}
0 = R(r) = 
- \big( K+f_r(r) \big) \Delta  +
 \Big( (r^2+a^2) \omega _0 + a p_{\varphi} \Big) ^2
\, , 
\end{equation}
\begin{gather}
\nonumber
0 = R'(r) = 
- \big( K+f_r(r) \big) 2(r-m) 
\\
\label{eq:Rp}
+
\dfrac{4rp_t}{c} 
\Big( (r^2+a^2) \omega _0 + a p_{\varphi} \Big) 
- f_r '(r) \, \Delta 
\, .
\end{gather}
These two equations can be solved for the constants of
motion $a p_{\varphi}$ and $K$,
\begin{equation}\label{eq:pphispher} 
a p_{\varphi}  = \, \dfrac{\omega _0}{(r-m)} 
\left( m(a^2-r^2) \pm 
r \Delta \sqrt{1-f_r'(r) \dfrac{(r-m)}{2 r^2 \omega _0 ^2}} \right)
\, ,
\end{equation}
\begin{equation}\label{eq:Kspher} 
K = \dfrac{r^2 \Delta \omega _0 ^2}{(r-m)^2}
\left(1 \pm \sqrt{1-f_r'(r) \dfrac{(r-m)}{2 r^2 \omega _0 ^2}} \right) ^2
-f_r(r) \, .
\end{equation}
As the left-hand side of (\ref{eq:dottheta}) is the square
of a real quantity, the right-hand side must be non-negative,
\begin{equation}\label{eq:dottheta2}
0 \le K 
-\Big( \dfrac{p_{\varphi}}{\mathrm{sin} \, \vartheta} +
a \,  \mathrm{sin} \, \vartheta \, \omega _0  \Big)^2
- f_{\vartheta} ( \vartheta ) 
\, ,  
\end{equation}
hence
\begin{equation}\label{eq:dottheta3}
\Big( K -f_{\vartheta} ( \vartheta ) \Big) a^2
\mathrm{sin} ^2 \vartheta \ge 
\left( a p_{\varphi} +
a^2 \mathrm{sin} ^2 \vartheta \, \omega _0 \right) ^2
\, .  
\end{equation}
Inserting (\ref{eq:pphispher}) and (\ref{eq:Kspher})
into (\ref{eq:dottheta3}) gives the photon region,
\begin{widetext}
\begin{gather}\label{eq:photonreg}
\left( \dfrac{r^2 \Delta }{(r-m)^2}
\left(1 \pm \sqrt{1-f_r'(r) \dfrac{(r-m)}{2 r^2 \omega _0^2}} \right)^2
- \frac{f_r(r)+ f_{\vartheta} ( \vartheta )}{\omega _0 ^2} \right) a^2
\mathrm{sin} ^2 \vartheta \ge 
\\
\nonumber
\left( \dfrac{1}{(r-m)} 
\left( m(a^2-r^2) \pm 
r \Delta \sqrt{1-f_r'(r) \dfrac{(r-m)}{2 r^2 \omega _0 ^2}} \right)+ 
a^2 \mathrm{sin} ^2 \vartheta  \right) ^2
\, .  
\end{gather}
\end{widetext}
Through each point with coordinates $(r , \vartheta)$ where this inequality
is satisfied, either with the plus or with the minus sign, there is a spherical
light ray.
 
In general, the photon region consists of stable and of unstable
spherical light rays. For an unstable spherical light ray one must have
\begin{gather} 
\nonumber
0 < R''(r) = 
- 2 K - 2 f_r(r) +
4 \, \omega _0 
\Big( (r^2+a^2) \omega _0 + a p_{\varphi} \Big) 
\\
\label{eq:Rpp}
+
8 \, r^2 \omega _0 ^2 
- f_r '(r) \, 4 (r-m) -f_r''(r) \Delta
\end{gather}
where $a \, p _{\varphi}$ and $K$ have to be expressed in
terms of the radius $r$ and the frequency $\omega _0$ with
the help of (\ref{eq:pphispher}) and (\ref{eq:Kspher}).

%---------------------------------------------------------------------
\section{Shadow of a Kerr black hole in a plasma}\label{sec:shadow}

For constructing the shadow, we fix an 
observer at Boyer-Lindquist coordinates
$\big( r_{\mathrm{O}}, \vartheta_{\mathrm{O}}
\big)$ with $r_{\mathrm{O}} > m + \sqrt{m^2-a^2}$
and a tetrad
\begin{gather}
\nonumber
	e_{0} = \left.
		\frac{(r^2+a^2)  \, \partial_t + a \, c \, \partial_{\varphi}}{
		c \, \rho \, \sqrt{\Delta}}
		\right|_{(r_{\mathrm{O}},\vartheta_{\mathrm{O}})}  , \quad
\\
\nonumber
	e_{1} = \left.
		\dfrac{1}{\rho} \, \partial_{\vartheta}
		\right|_{(r_{\mathrm{O}},\vartheta_{\mathrm{O}})} , 
\\
\nonumber
	e_{2} = \left.
		\frac{-\partial_{\varphi} - a \mathrm{sin} ^2 \vartheta \partial_t
		}{
		\rho  \mathrm{sin} \, \vartheta} 
		\right|_{(r_{\mathrm{O}},\vartheta_{\mathrm{O}})} ,
\\
\label{eq:tetrad}
	e_{3} = \left.
		-\dfrac{\sqrt{\Delta}}{\rho} \, \partial_r
		\right|_{(r_{\mathrm{O}},\vartheta_{\mathrm{O}})}  .
\end{gather}
This tetrad is well-defined and orthonormal for any observer position
in the domain of outer communication. It is chosen such that the ingoing 
and outgoing principal null directions of the Kerr spacetime are in the plane
spanned by $e_0$ and $e_3$. 
We construct the shadow for the case that $c \, e_0$ is the 
4-velocity of the observer, following closely the 
procedure of Grenzebach et al. 
\cite{GrenzebachPerlickLaemmerzahl2014}.
 
We consider light rays issuing from the observer position
into the past. For each light ray $\lambda (s)$ with coordinate representation
$\bigl(r(s),\vartheta(s),\varphi(s),t(s)\bigr)$, we write the tangent
vector as
\begin{equation}\label{eq:dotlambda1}
	\dot{\lambda} = \dot{r} \partial_{r} + \dot{\vartheta} \partial_{\vartheta}
	+ \dot{\varphi} \partial_{\varphi} + \dot{t} \partial_{t}
\end{equation}
where an overdot means derivative with respect to the curve parameter $s$.
On the other hand, the tangent vector at the observation event
can be written as
\begin{equation}\label{eq:dotlambda2}
	\dot{\lambda} = - \alpha  \, e_0 + \beta \Big( \mathrm{sin} \,  \theta \, 
\mathrm{cos} \, \psi \, e_1  
	+ \mathrm{sin} \, \theta \, \mathrm{sin} \, \psi \, e_{2} 
+ \mathrm{cos} \, \theta \, e_3 \Big) 
\end{equation}
where $\alpha$ and $\beta$ are positive factors. Recall that, by (\ref{eq:Hg}),
we have parametrized the light rays such that 
$g \big( \dot{\lambda} , \dot{ \lambda} \big) = - \omega _p ^2$. 
Therefore,    $\alpha$ and $\beta$ must be related by
\begin{equation}\label{eq:alphabeta}
\alpha ^2 - \beta ^2 = \omega _p ^2 \big| _{(r_{\mathrm{O}}, \vartheta _{\mathrm{O}})}
\, .
\end{equation}
Eq.  (\ref{eq:dotlambda2}) 
defines the celestial coordinates $\theta$ and $\psi$ for our observer.  
$\theta$ is the colatitude and $\psi$ is the azimuthal angle. The poles
$\theta =0$ and $\theta = \pi$ correspond, respectively, to ingoing and 
outgoing past-oriented principal null rays. In this sense, we may say that
$\theta =0$ is ``the direction towards the black hole'' and $\theta = \pi$
is ``the direction away from the black hole'' on the observer's sky.

For each light ray, $\alpha$ is determined by 
\begin{gather}\label{eq:alpha}
	\alpha = 
		g \big(\dot{\lambda},e_0 \big) =
		\dfrac{1}{\rho \sqrt{\Delta}}
g \Big( \dot{\lambda} , (r^2+a^2) \dfrac{1}{c}\partial _t + a  \partial _{\varphi} \Big) 
\\
\nonumber
= 
\dfrac{(r^2+a^2)}{c \rho \sqrt{\Delta}} \big( \dot{t} g_{tt} + \dot{\varphi} g_{\varphi t} \big)
+ \dfrac{a}{\rho \sqrt{\Delta}} \big( \dot{t} g_{t \varphi} + \dot{\varphi} g_{\varphi \varphi} \big)
\\
\nonumber
= 
\,  \dfrac{(r^2+a^2)}{\rho \sqrt{\Delta}} \omega _0
+ \dfrac{a}{\rho \sqrt{\Delta}} p_{ \varphi}
\end{gather}
hence
\begin{equation}\label{eq:beta}
\beta = \sqrt{ 
\dfrac{1}{\rho ^2 \Delta}  \Big( \big( r^2+a^2 \big) \, \omega _0  
+ a p_{\varphi} \Big)^2
- \omega _p ^2} 
\end{equation}
where all expressions are to be evaluated at $(r_{\mathrm{O}}, \vartheta _{\mathrm{O}})$.
We will now determine how the constants of motion $p_{\varphi}$ and $K$ of a 
light ray are related to the celestial coordinates $\theta$ and $\psi$. To that end
we
compare coefficients of $\partial _r$ in (\ref{eq:dotlambda1}) and (\ref{eq:dotlambda2})
which yields
\begin{equation}\label{eq:cc1}
- \beta \, \mathrm{cos} \, \theta \, \dfrac{\sqrt{\Delta}}{\rho} = \dot{r}
\, .
\end{equation}
Upon squaring both sides we find, with the help of  (\ref{eq:dotr}) and
(\ref{eq:beta}),
\begin{gather}
\left( \Big((r^2+a^2) \, \omega _0 +a \, p _{\varphi} \Big)^2
- \Delta \rho ^2 \omega _p ^2 \right) \big( 1 - \mathrm{sin} ^2 \theta \big)
\nonumber
\\
\label{eq:cc2}
= -\big(K+ f_r (r) \big) \Delta + \Big((r^2+a^2) \, \omega _0 
+ a \, p_{\varphi} \Big)^2
\, .
\end{gather}
Solving for $\mathrm{sin} ^2 \theta$ and taking the square root (using $\mathrm{sin} \, 
\theta \ge 0$ as $0 \le \theta \le \pi$) results in
\begin{gather}\label{eq:shtheta}
\mathrm{sin} \, \theta = 
\\
\nonumber
\left.
\sqrt{\dfrac{\big(K-f_{\vartheta} ( \vartheta ) \big) \Delta}{
 \Big( (r^2+a^2)\, \omega _0 + a \, p_{\varphi} \Big)^2
- \Big( f_r(r)+f_{\vartheta} ( \vartheta ) \Big) \Delta }}
		\right|_{(r_{\mathrm{O}},\vartheta_{\mathrm{O}})}  \, .
\end{gather}
Similarly, comparing coefficients of $\partial _{\varphi}$ in 
(\ref{eq:dotlambda1}) and (\ref{eq:dotlambda2}) yields
\begin{equation}\label{eq:cc3}
- \, \dfrac{ \alpha \, a}{\rho \sqrt{\Delta}} \, - \,
\dfrac{\beta \, \mathrm{sin} \, \theta \, \mathrm{sin} \, \psi}{
\rho \, \mathrm{sin} \, \vartheta}
\, = \, \dot{\varphi} \, .
\end{equation}
Upon inserting (\ref{eq:dotphi}), (\ref{eq:alpha}),
(\ref{eq:beta}) and (\ref{eq:shtheta}) into 
(\ref{eq:cc3}) we find
\begin{equation}\label{eq:shpsi}
\mathrm{sin} \, \psi = \left.
\dfrac{ - p_{\varphi} - a \, \mathrm{sin} ^2 \vartheta \, \omega _0 }{
\mathrm{sin} \, \vartheta \, \sqrt{K-f_{\vartheta} ( \vartheta )}}
	\right|_{\vartheta_{\mathrm{O}}}  \, .
\end{equation}
The shadow is the set of all points on the observer's sky, i.e.,
of tangent vectors to light rays, such that past-oriented light rays
with such a tangent vector go to the horizon.
Light rays that correspond to boundary points of the shadow spiral
asymptotically towards spherical light rays, so they must have the 
same values for $K$ and $p_{\varphi}$ as these limiting spherical
light rays. If $a \neq 0$, we can insert these values for $p_{\varphi}$ and
$K$ from (\ref{eq:pphispher}) and (\ref{eq:Kspher}), respectively, where 
$r=r_p$ runs over the radius values of the limiting spherical light rays.
With these expressions for $K(r_p)$ and $p_{\varphi}(r_p)$ inserted
into (\ref{eq:shtheta}) and (\ref{eq:shpsi}) we get the boundary 
of the shadow  as a curve on the observer's sky parametrized by $r_p$.
The boundary curve consists of a lower part, where $\psi$ runs from 
$- \pi/2$ to $\pi /2$, and an upper part where $\psi$ runs from $\pi /2$
to $3 \pi /2$. The parameter $r_p$ runs from a minimum value 
$r_{p, \mathrm{min}}$ to a maximum value $r_{p, \mathrm{max}}$
and then back to $r_{p, \mathrm{min}}$. The values $r_{p, \mathrm{min}}$  
and $r_{p, \mathrm{max}}$ are determined by the property that then
$\mathrm{sin} ^2 \psi$ must be equal to 1, i.e., by (\ref{eq:shpsi}),
$r_{p} = r_{p, \mathrm{min}/\mathrm{max}}$ if
\begin{equation}\label{eq:rpminmax}
- p_{\varphi} (r_p ) - \, a \, \mathrm{sin} ^2 \vartheta _{\mathrm{O}}
\, \omega _0
 \, = \, \pm \, 
\mathrm{sin} \, \vartheta _{\mathrm{O}}
 \, \sqrt{K(r_p)-f_{\vartheta } ( \vartheta _{\mathrm{O}})}
\, .
\end{equation}
Comparison with (\ref{eq:photonreg}) shows that 
$ r_{p, \mathrm{min}/\mathrm{max}}$ are the radius
values of spherical light rays that have turning points
at $\vartheta = \vartheta _{\mathrm{O}}$. In other words,
we get the interval of allowed $r_p$ values by intersecting
the photon region with the cone  $\vartheta = \vartheta _{\mathrm{O}}$.
Each value of $r_p$ in the interval $\,] \, r_{p, \mathrm{min}} , 
r_{p, \mathrm{max}} \, [ \,$ corresponds to two points on
the boundary curve of the shadow whose $\psi$ coordinates
$\psi _1 \in \; ] - \pi /2, \pi /2 \, [ \,$ and $\psi _2 \in \; ] \, \pi/2
, 3 \pi /2 \, [ \,$ are related by $\mathrm{sin} \, \psi _1 = 
\mathrm{sin} \, \psi_2$.
The corresponding $\theta$ coordinates are the same,
$\theta _1 = \theta _2$, because according to (\ref{eq:shtheta}) 
they are uniquely determined by $K (r_p)$ and $p_{\varphi} (r_p)$.
This demonstrates the 
remarkable fact that the shadow is always symmetric with respect 
to a horizontal axis. This symmetry was not to be expected for an 
observer off the equatorial plane and a plasma density
depending on $\vartheta$.     

Our equations (\ref{eq:shtheta}) and (\ref{eq:shpsi}) give us the
shape and the size of the shadow for an observer anywhere in the domain
of outer communication, providing that the observer's four-velocity is
proportional to our basis vector $e_0$. If the observer is moving
with a different four-velocity, the shadow is distorted by aberration,
see Grenzebach \cite{Grenzebach2015}. When plotting the shadow, below
in Figs.~\ref{fig:e1}, \ref{fig:e2}, \ref{fig:e3} and \ref{fig:e4sh}, we use
stereographic projection onto a plane that is tangent to the celestial
sphere at the pole $\theta =0$, and in this plane we use (dimensionless)
Cartesian coordinates,
\[
	X(r_p) = -2 \tan \Big( \frac{\theta(r_p)}{2} \Big)
		\sin \big( \psi(r_p) \big) \, ,
\]
\begin{equation}\label{eq:stereo}
Y(r_p) = -2 \tan \Big( \frac{\theta(r_p)}{2} \Big)
		\cos \big( \psi(r_p) \big).
\end{equation}
We indicate these Cartesian coordinate axes by cross-hairs in our
plots. 
Recall that the pole $\theta =0$, i.e., the origin of our Cartesian 
coordinate system, corresponds to a past-oriented ingoing
principal null ray through the observer position.
This method of plotting the shadow, which was also used in
\cite{GrenzebachPerlickLaemmerzahl2014}, is to be distinguished
from the one introduced by Bardeen \cite{Bardeen1973}. Firstly,
Bardeen considers an observer at infinity while we allow  any
observer position in the domain of outer communication. Secondly,
Bardeen plots the shadow on a plane where (dimensionful) impact
parameters are used as the coordinates on the axes, while our
(dimensionless) coordinates are directly related to 
angular measures on the observer's sky. Thirdly, the origin of 
Bardeen's coordinates
corresponds to a null ray with $p_{\varphi} =0$ while our origin
corresponds to a principal null ray which has $p_{\varphi} =-a \sin \vartheta_{\textrm{O}}$;
of course, the choice of the origin is a matter of convention. One just has to keep in mind that our way of plotting
may be directly compared with Bardeen's only if the observer is at a big radius
coordinate and that the origin is horizontally shifted.
  
For the construction of the shadow we prescribe the 
constant of motion $\omega_0$. There are two different situations 
that may be considered. (a) Firstly, we may think of static light sources 
distributed at big radius coordinates that emit light rays monochromatically 
with this frequency $\omega _0$. Note that these light rays
arrive at the observer, whose four-velocity $e_0 c$ is given 
by (\ref{eq:tetrad}), with different frequencies, i.e., in general
this observer will see the shadow against a backdrop that is not 
monochromatic. This is true also for the shadow without a plasma. 
Only if the observer is static and far away from the black hole will 
the backdrop be monochromatic but, for any position of the observer,
we may directly use our formula for the boundary curve of the shadow
with the prescribed $\omega _0$. (b) Secondly, we may think 
of light sources that emit a wide range of frequencies, and of an observer 
that filters out a particular frequency $\omega _{\mathrm{obs}}$. Then we 
have to express $\omega _0$ in our equations for the shadow in terms of 
$\omega _{\mathrm{obs}}$. The desired relation between the two frequencies
follows from (\ref{eq:freq}) with $U^{\mu}=e^{\mu}_0 c$ if $e_0^{\mu}$ is
inserted from (\ref{eq:tetrad}). We find that
\begin{equation}\label{eq:omegaobs}
\omega _{\mathrm{obs}} = 
\dfrac{(r^2+a^2) \omega _0+a \, p_{\varphi}}{\rho \sqrt{\Delta}}
\Bigg| _{\big( r_{\mathrm{O}},\vartheta _{\mathrm{O}} \big)} \, .
\end{equation}
After solving this equation for $\omega _0$, we are able to replace 
$\omega _0$ in (\ref{eq:shtheta}) and (\ref{eq:shpsi}) by 
$\omega _{\mathrm{obs}}$. The same substitution has to be made 
in (\ref{eq:pphispher}) and (\ref{eq:Kspher}) with $r= r_p$; 
inserting the latter expressions into (\ref{eq:shtheta}) and (\ref{eq:shpsi})
gives the boundary curve of the shadow parametrized by $r_p$, now for prescribed
$\omega _{\mathrm{obs}}$. The range of the curve parameter $r_p$ has to be
determined for each value of $\omega _{\mathrm{obs}}$ individually. Note that,
by (\ref{eq:omegaobs}),  $\omega _{\mathrm{obs}} \to \omega _0$ for $r_{\mathrm{O}}
\to \infty$, i.e., that there is no difference between case (a) and case (b) if
the observer is at a big radius coordinate.

For the reader's convenience, we end this section by summarizing the
construction method of the shadow in a step-by-step procedure.

1. Choose the mass parameter $m$ and the spin parameter $a$ with $a^2 \le m^2$. 
The mass parameter $m$ gives a natural length unit, i.e., all other lengths may be given
in units of $m$.

2. Choose a plasma frequency $\omega _p ( r , \vartheta )$ around the black hole 
which is related to the electron number density by the formula (\ref{eq:omega_p}).
The plasma frequency has to satisfy the separability condition (\ref{eq:sepcon}), so 
the plasma distribution is characterized by two functions $f_r(r)$ and 
$f_{\vartheta} ( \vartheta )$. The refractive index of the plasma 
is then given by the expression (\ref{eq:n}). Only for such plasma frequencies 
are the ray equations completely integrable and the shadow can be calculated 
analytically. 

3. Choose a position of an observer anywhere in the domain of outer communication 
by prescribing its radial and angular coordinate $r_\mathrm{O}$ and $\vartheta_\mathrm{O}$. 
For an illustration see Fig. 7 in \cite{GrenzebachPerlickLaemmerzahl2014}.

4. Choose the constant of motion $\omega_0$ for the rays that are to be considered. 
%Note 
%that $\omega_0$ is the photon frequency measured by an observer on a $t$-line at infinity.
%If the observer is not far away from the black hole, or not on a $t$-line, the observed
%frequency may be considerably different from $\omega _0$. The observed frequency is given
%by (\ref{eq:freq}) where $U^{\mu} (x)$ is the four-velocity of the observer. 
In the formulas for the shadow, $\omega _0$ will enter only in terms of the quotient 
$\omega _p (r , \vartheta)^2/\omega _0^2$. Therefore, it is convenient to give 
$\omega _p (r , \vartheta )$ and $\omega _0$ as multiples of the same frequency
unit $\omega _c$ which will then drop out from all relevant formulas. 

5. Write the celestial coordinates $\sin \theta$ and $\sin \psi$ in terms of the 
constants of motion of the corresponding ray by  
(\ref{eq:shtheta}) and (\ref{eq:shpsi}) with $r=r_\mathrm{O}$ and 
$\vartheta=\vartheta_\mathrm{O}$ substituted. For an illustration of the 
angles $\theta$ and $\psi$ see Fig. 8 in \cite{GrenzebachPerlickLaemmerzahl2014}.

6. Substitute into these expressions for $\sin \theta$ and $\sin \psi$
the expressions $K(r_p)$ and $p_\varphi(r_p)$ according to formulas (\ref{eq:pphispher})  
and (\ref{eq:Kspher}) with $r=r_p$. Here $r_p$ runs over an interval of radius 
coordinates for which unstable spherical light rays exist. This gives us $\sin \theta$ and 
$\sin \psi$ as functions of  $r_p$, i.e., it gives us a curve on the observer's
sky. This is the boundary curve of the shadow. In the next two steps we determine
the range of the curve parameter $r_p$. (Here we assume $a \neq 0$. In the 
non-spinning case $a=0$ the shadow is circular  
and the equation for $\sin \theta$ gives us directly the angular radius of the shadow.) Note that, by (\ref{eq:pphispher}) and (\ref{eq:Kspher}), there is a sign
ambiguity in the expressions for $K(r_p)$ and $p_\varphi(r_p)$. We will show below that
for a wide range of plasma distributions, including all the examples considered in this
paper, the equations can hold only with the plus sign. However, it is easy to construct 
mathematical examples where solutions with the minus sign occur.

7. Solve the equation 
$\sin \psi(r_p) = 1$ for $r_p$. This gives us the minimal value $r_{p, \mathrm{min}}$. 
Note that there could be several solutions of this equation
which are real and $>m$. Then one has to determine, depending on the position
of the light sources and of the observer, which one is relevant for the 
formation of the shadow. E.g., if there are two photon regions with unstable
spherical light rays and if the observer is outside the outer one, there
are two possible values for $r_{p,\mathrm{min}}$: If we stick with our general rule
that there are light sources everywhere but not between the observer and the black
hole (i.e., not in the region crossed by past-oriented light rays that approach the
horizon), then we have to choose for $r_{p,\mathrm{min}}$ a value on the boundary
of the inner photon region. However, if there are light sources only at big radius
values, then we have to choose a value on the inner boundary of the outer photon region.
If the plasma density is small and if
the observer is in the equatorial plane, for small 
$a$ it will be $r_{p, \mathrm{min}} \lesssim 3m$, while for a nearly extreme Kerr 
black hole ($a \lesssim m$) it will be $r_{p, \mathrm{min}} \gtrsim m$.

8. Solve the equation $\sin \psi(r_p) = -1$ for $r_p$. This gives us the maximal
value $r_{p, \mathrm{max}}$. As in the case of $r_{p,\mathrm{min}}$, 
there may be several solutions. If the plasma density is small,
for small $a$ it will be 
$r_{p, \mathrm{max}} \gtrsim 3m$, while for a nearly extreme Kerr black hole it 
will be $r_{p, \mathrm{max}} \lesssim 4m$.

9. Calculate $\sin \theta(r_p)$ and $\sin \psi(r_p)$ where $r_p$ ranges over the  
interval $\,] \, r_{p, \mathrm{min}} , r_{p, \mathrm{max}} \, [ \,$. 
Note that $\theta(r_{p, \mathrm{min}}) +
\theta(r_{p, \mathrm{max}})$ gives the horizontal diameter of the shadow.

10. Calculate the dimensionless Cartesian coordinates $X$ and $Y$ of the boundary curve
of the shadow by formulas (\ref{eq:stereo}). Choosing $-\pi /2 \le \psi (r_p) \le \pi /2$
and letting $r_p$ run from $r_{p, \mathrm{min}}$ to $r_{p, \mathrm{max}}$ gives the 
lower half of the boundary curve of the shadow. The upper half of the curve is the
mirror image of the lower half with respect to a horizontal axis.

11. The preceding instruction applies to the case that $\omega _0$ is prescribed and 
that the observer has four-velocity $e_0$ as given in (\ref{eq:tetrad}).
If the shadow is to be determined for prescribed $\omega _{\mathrm{obs}}$, one has 
to express $\omega _0$ in terms of $\omega _{\mathrm{obs}}$ by (\ref{eq:omegaobs}),
as outlined above. If the shadow is to be determined for an observer 
whose four-velocity is not equal to $e_0$, one
has to apply the special-relativistic aberration formula, cf. \cite{Grenzebach2015} 
(and, in the case
that $\omega _{\mathrm{obs}}$ is prescribed, the special-relativistic Doppler
formula for $\omega _{\mathrm{obs}}$).

%------------------------------------------------------------------------ 
\section{The case of a low density plasma}\label{sec:approx}

In this section we consider the case that the plasma frequency is small
in comparison to the photon frequency. More precisely, we assume that
the separability condition (\ref{eq:sepcon}) is satisfied and that the
functions $f_r(r)$, $f_{\vartheta} ( \vartheta )$, $f_r'(r)$ and 
$f_{\vartheta}' (\vartheta)$ are so small that all expressions can be
linearized with respect to these functions. (Actually, the function
$f_{\vartheta} ' (\vartheta )$ will not occur in the following
calculations.) 

We will first show that in this case (\ref{eq:Kspher}) and, thus, 
(\ref{eq:pphispher}) can hold only with the plus sign. By contradiction,
let us assume that (\ref{eq:Kspher}) holds with the minus sign.
Then linearization of the square-root in (\ref{eq:Kspher})
yields 
\begin{equation}\label{eq:Klin1}
K =  - f_r(r) + \, \dots \, , 
\end{equation}
hence
\begin{equation}\label{eq:Klin2}
K - f_{\vartheta} ( \vartheta ) = - \omega _p ( r , \vartheta ) ^2 \rho ^2 + 
\, \dots , .
\end{equation}
This is a contradiction because, by (\ref{eq:ptheta}), 
the left-hand side of (\ref{eq:Klin2}) cannot be negative.

We are now ready to calculate a linear correction to the shadow due to the presence
of a low density plasma. The boundary curve of the shadow for the observer at 
($r_{\mathrm{O}}$,$\vartheta_{\mathrm{O}}$) is determined by equations 
(\ref{eq:shtheta}) and (\ref{eq:shpsi}) with $K(r_p)$ and $p_{\varphi}(r_p)$ given by
(\ref{eq:Kspher}) and (\ref{eq:pphispher}) with the plus signs. Here $r_p$ runs over 
the radius values of spherical light rays which satisfy (\ref{eq:photonreg}) with the 
plus sign. This gives us the angles $\theta$ and $\psi$ as functions of $r_p$, 
for a given observer position ($r_{\mathrm{O}}$,$\vartheta_{\mathrm{O}}$).

\begin{widetext}

Expanding (\ref{eq:shtheta}) and (\ref{eq:shpsi}), we obtain:
\begin{equation}\label{small-theta}
\sin \theta(r_p) = \frac{ 2r_p \sqrt{\Delta _p \Delta_\mathrm{O}} }{|Z|} \, + \, 
\frac{(r_p-m)\sqrt{\Delta _p \Delta_\mathrm{O}} \, 
[\Delta _p f_r'(r_p) Z + 2 \Delta_\mathrm{O} (f_r(r_\mathrm{O}) 
+ f_\vartheta(\vartheta_\mathrm{O})) r_p(r_p-m)  ]   }{2\omega _0^2 |Z|^3} \, -
\end{equation}
$$
- \, \frac{(r_p-m) \sqrt{\Delta _p \Delta_\mathrm{O}} \, 
[\Delta _p f_r'(r_p) +f_r(r_p)(r_p-m)+ f_\vartheta(\vartheta_\mathrm{O}) (r_p-m)   ] 
}{4  r_p  \omega _0^2 \Delta _p |Z|} ,
$$

\begin{equation} \label{small-psi}
\sin \psi(r_p) = - \, \frac{C_1}{a \sqrt{\Delta _p } \sin \vartheta_\mathrm{O} r_p} 
\, - \,\frac{C_1 (r_p-m) \, [\Delta _p f_r'(r_p) +f_r(r_p)(r_p-m)+ 
 f_\vartheta(\vartheta_\mathrm{O}) (r_p-m)   ]  
 }{ 16 \omega _0^2 a \Delta _p ^{3/2} \sin \vartheta_\mathrm{O} r_p^3} \, + \,  
 \frac{\Delta _p f_r'(r_p) (r_p-m)}{8\omega _0^2 a r_p^2 \sin \vartheta_\mathrm{O}} ,
\end{equation}

\end{widetext}
where
\begin{equation}
Z = r_\mathrm{O}^2(r_p-m) + a^2 r_p -mr_p^2 + r_p \Delta ,
\end{equation}
\begin{equation}
C_1 = a^2 \sin^2 \vartheta_\mathrm{O} (r_p-m) + m(a^2-r_p^2) + r_p\Delta ,
\end{equation}
\begin{equation}
\Delta _p = r_p^2 +a^2 - 2mr_p , \quad 
\Delta_\mathrm{O} = r_\mathrm{O}^2 + a^2 -2mr_\mathrm{O}.
\end{equation}
Here $r_p$ is varying between a minimum value $r_{p, \mathrm{min}}$ and a maximum value 
$r_{p, \mathrm{max}}$ which are found from the equations $\sin \psi(r_p) = 1$ and 
$\sin \psi(r_p) = -1$, respectively. 

In formulas (\ref{small-theta}) and (\ref{small-psi}) the first terms are vacuum terms, 
while the other terms are proportional to $1/\omega_0^2$. So in the case of high 
frequencies the plasma terms tend to zero.

%-----------------------------------------------------------------------------------
\section{The Schwarzschild case}\label{sec:Schwarzschild}

In the Schwarzschild case, $a=0$, the separability
condition (\ref{eq:sepcon}) requires the plasma 
frequency to be of the form
\begin{equation}\label{eq:sepconS}
\omega _p (r , \vartheta ) ^2 = 
\dfrac{f_r(r)+f_{\vartheta} ( \vartheta )}{r^2}
\, .
\end{equation}
This condition is, of course, in particular satisfied
if $\omega _p$ depends only on $r$; this case was treated
in our earlier paper \cite{PerlickTsupkoBK2015}. However, 
the separability condition is also
satisfied for some $\vartheta$ dependent distributions
that may be considered, e.g.,  as reasonable models for a 
plasma that rotates about a Schwarzschild black hole.  In
the following we discuss this more general case. 

For $a=0$, the inequality (\ref{eq:photonreg}) reduces
to an equality, as the right-hand side cannot be negative,
so we get
\begin{equation}\label{eq:photsphere1}
0 = 
 \dfrac{\omega _0 ^2}{(r-m)^2} 
\left( -m r^2 \pm 
r^2 (r-2m) \sqrt{1-f_r'(r) \dfrac{(r-m)}{2r^2 \omega _0^2}} \right) ^2
\, .
\end{equation}
Hence, the photon region degenerates to a photon sphere
(or to several photon spheres) in this case.
As we restrict to the domain of outer communication where $r>2m$, only the 
upper sign is possible and (\ref{eq:photsphere1})
can be rewritten as
\begin{equation}\label{eq:photsphere2}
m = 
(r-2m) \sqrt{1-f_r'(r) \dfrac{(r-m)}{2r^2 \omega _0^2}} 
\, .  
\end{equation}
The same result follows from (\ref{eq:pphispher}). 
Upon squaring, the condition for a photon sphere can be
rewritten as
\begin{equation}\label{eq:photsphere3}
(r-3m) \, 2 \, \omega _0 ^2 = \Big( 1- \dfrac{2m}{r} \Big) ^2 f_r'(r)
\, .  
\end{equation}
By (\ref{eq:Kspher}) and (\ref{eq:photsphere2}), 
for a light ray on a photon sphere at $r=r_p$ 
the  Carter constant takes the value
\begin{equation}\label{eq:Krp}
K = \dfrac{r_p^3 \omega _0 ^2}{(r_p-2m)}-f_r(r_p)
\, . 
\end{equation}
The radius of the photon sphere and the corresponding
Carter constant depend on $f_r(r)$ and 
on the frequency $\omega _0$ but not, remarkably, on 
$f_{\vartheta} (\vartheta )$. If $f_r'(r)=0$, we can 
incorporate the constant $f_r(r)$ into the function
$f_{\vartheta} ( \vartheta )$, i.e., $\omega _p (r , \vartheta ) ^2 = 
f_{\vartheta} ( \vartheta ) /r^2$. Then  the photon sphere is at 
$r=3m$, for any choice of $f_{\vartheta} ( \vartheta )$ and all 
frequencies $\omega _0$, as can be read from (\ref{eq:photsphere3}). 

For the formation of a shadow we have to choose $f_r (r)$ and 
$f_{\vartheta} ( \vartheta )$ such that there is an unstable photon 
sphere at a radius $r=r_p$. (If there are several unstable photon spheres,
one has to determine which one is relevant for the shadow, depending
on the position of the observer.) For $a=0$ (\ref{eq:shtheta}) reduces to
\begin{equation}\label{eq:shthetaS}
\mathrm{sin} \, \theta = \left.
\sqrt{\dfrac{K-f_{\vartheta} ( \vartheta )}{
\dfrac{r^3 \omega _0^2}{(r-2m)} 
- f_r(r)-f_{\vartheta} ( \vartheta ) }}
		\right|_{(r_{\mathrm{O}},\vartheta_{\mathrm{O}})}  \, .
\end{equation}
Inserting into (\ref{eq:shthetaS})
the fixed value of $K$ from  (\ref{eq:Krp}) gives the angular radius
of the shadow which is circular,
\begin{equation}\label{eq:shthetaS2}
\mathrm{sin} \, \theta = 
\sqrt{\dfrac{r_p^3 (r_{\mathrm{O}} - 2 \, m )}{r_{\mathrm{O}}^3 (r_p - 2 \, m )}} \, \times
\end{equation}
\[
 \times  \, \sqrt{ 
\dfrac{
1 - \Big(f_r(r_p) + f_{\vartheta} ( \vartheta _{\mathrm{O}}) \Big) 
\dfrac{(r_p-2 \, m)}{r_p^3 \omega _0 ^2}
}{
1 - \Big(f_r(r_{\mathrm{O}}) + f_{\vartheta} ( \vartheta _{\mathrm{O}}) \Big) 
\dfrac{(r_{\mathrm{O}}-2 \, m)}{r_{\mathrm{O}}^3 \omega _0 ^2}
}
}
\, .
\]
In the Schwarzschild case Eq. (\ref{eq:shpsi}) gives the relation between the 
azimuthal angle $\psi$ and the constant of motion $p_{\varphi}$
but no further information on the shadow. From (\ref{eq:shthetaS2}) we read
that the plasma has an increasing effect of the shadow if and only if
\begin{equation}\label{eq:increase}
\omega _p \big( r_{\mathrm{O}}, \vartheta _{\mathrm{O}} \big) ^2
\Big( 1- \frac{2 \, m}{r_{\mathrm{O}}} \Big) >
\omega _p \big( r_p , \vartheta _{\mathrm{O}} \big) ^2
\Big( 1- \frac{2 \, m}{r_p} \Big) \, .
\end{equation}
This inequality is necessarily violated if the observer is in a vacuum 
region, i.e., if $\omega _p \big( r_{\mathrm{O}}, \vartheta _{\mathrm{O}} \big)
=0$. Moreover, if $\omega _p \big( r_{\mathrm{O}}, \vartheta _{\mathrm{O}} \big)$
goes to zero for $r_{\mathrm{O}} \to \infty$, the inequality (\ref{eq:increase})
is necessarily violated for observers who are sufficiently far away from the
black hole. So in both cases the plasma either decreases the angular radius
of the shadow or leaves it unchanged. 

We summarize the 
results for a plasma with density $\omega _p (r , \vartheta ) ^2
= \big( f_r(r) + f_{\vartheta} ( \vartheta ) \big) /r^2$ on the 
Schwarzschild spacetime: There is no photon region but a photon
sphere (or several photon spheres), and the shadow is always circular,
although the plasma distribution is not necessarily spherically symmetric. The 
violation of spherical symmetry is reflected by the fact that,
by (\ref{eq:shthetaS2}), the angular
radius of the shadow depends on the $\vartheta$ coordinate of the
observer.
 
We want to further specify the equations for the shadow in a plasma
on the Schwarzschild spacetime for the case of a low density distribution. As
before, by that we mean that all expressions may be linearized with respect to 
$f_r(r)$, $f _{\vartheta} ( \vartheta )$ and $f_r' (r)$. 

Linearization of (\ref{eq:photsphere3}) yields
\begin{equation}\label{eq:rpSlin}
r_p = 3m + \frac{f'_r(3m)}{18\omega _0^2} \, \dots 
\end{equation}
As $f_{\vartheta} ( \vartheta )$ has no influence on the position of the photon sphere,
this result is in agreement with Eq.(33) of our earlier paper \cite{PerlickTsupkoBK2015}
where we considered the case $f_{\vartheta} ( \vartheta ) = 0$. 
Using (\ref{eq:rpSlin}), linearization of (\ref{eq:shthetaS2}) yields
\[
\sin \theta = \frac{3\sqrt{3} m \sqrt{r_\mathrm{O}-2m}}{r_\mathrm{O}^{3/2}} 
\Bigg( 1 + \frac{\omega _p \big( r_{\mathrm{O}} , \vartheta _{\mathrm{O}} )}{2 \,  \omega _0^2}
\Big( 1 - \frac{2 \, m}{r_{\mathrm{O}}} \Big)  \, -
\]
\begin{equation}\label{thetaSlin}
- \, \frac{\omega _p (3m, \vartheta _{\mathrm{O}} )^2}{6 \, \omega _0^2} + 
\, \dots \, \Bigg) .
\end{equation}
The condition (\ref{eq:increase}) for an increasing effect of the plasma on the 
shadow reads in the linear approximation
\begin{equation}
3 \, \omega _p \big( r_{\mathrm{O}} , \vartheta _{\mathrm{O}} \big) ^2
\, \Big( 1 - \frac{2 \, m}{r_{\mathrm{O}}} \Big) > 
\omega _p \big( 3m , \vartheta _{\mathrm{O}} \big) ^2 + \, \dots 
\end{equation}
We conclude this section by briefly discussing this condition for the case 
that the plasma frequency is independent of $\vartheta$ and depends on 
$r$ via a power law,
\begin{equation}
\frac{\omega_p (r)^2}{\omega_0^2} = \beta_0 \frac{m^k}{r^k} \, , \quad 
k \ge 0 \, , \quad 
\beta_0 = \text{constant} 
\end{equation}
which includes the case of a homogeneous plasma, with $k=0$.

\begin{figure}
\includegraphics[width=0.45\textwidth]{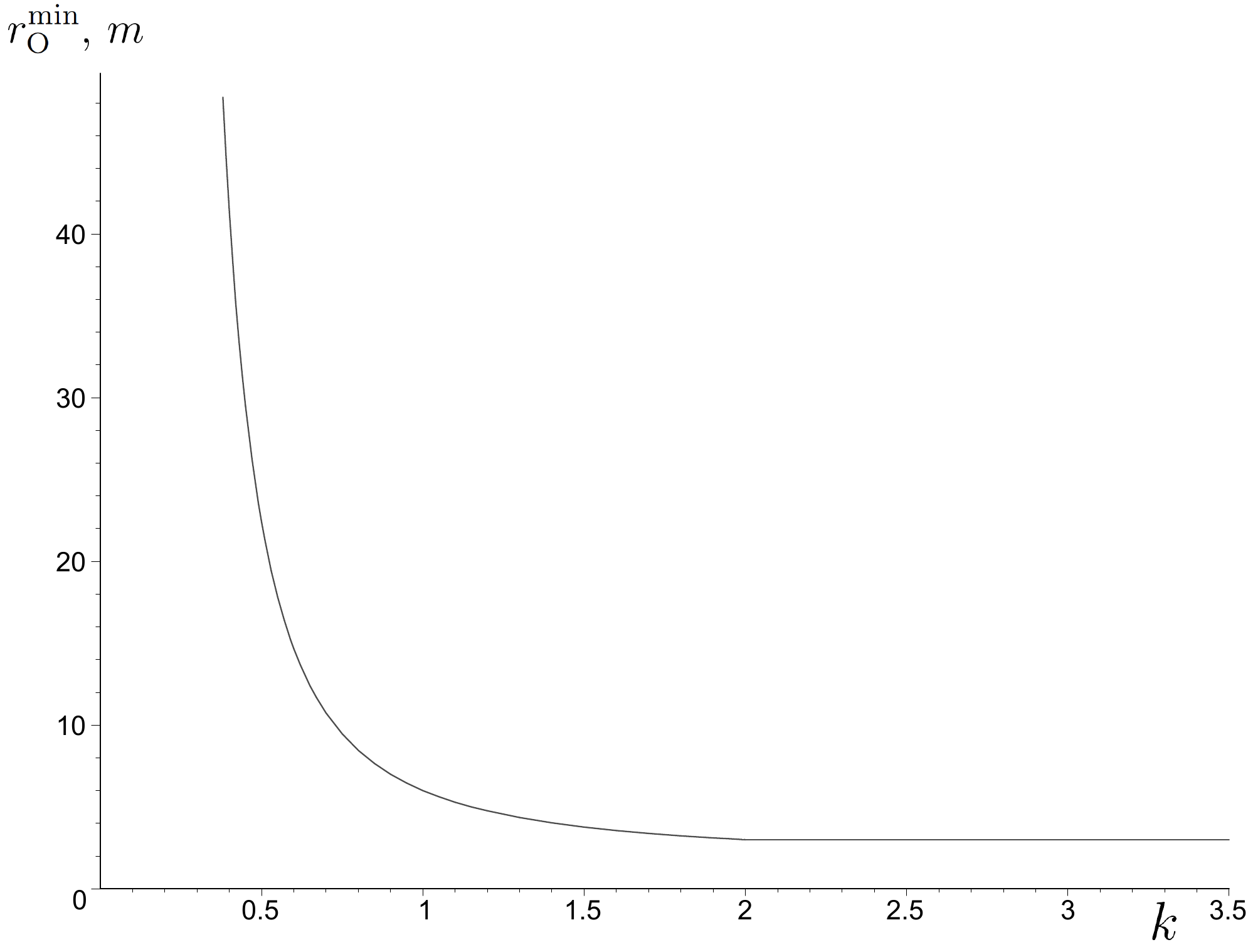}
%\centerline{\epsfig{figure=r0k.eps, width=0.45\textwidth}}
\caption{Dependence of $r_\mathrm{O}^{\mathrm{min}}(k)$ on $k$. The effect of the 
plasma on the shadow is increasing for an observer position 
$3m < r_\mathrm{O} < r_\mathrm{O}^{\mathrm{min}}(k)$ while it is decreasing
for $r_\mathrm{O} > r_\mathrm{O}^{\mathrm{min}}(k)$. For $k \ge 2$ we have
$r_\mathrm{O}^{\mathrm{min}}(k)=3m$, whereas for $k=1$ and $k=0$ we have   
$r_\mathrm{O}^{\mathrm{min}}(1)=6m$ and  $r_\mathrm{O}^{\mathrm{min}}(0)=\infty$, 
respectively. So in the inhomogeneous case $k>0$ an observer who is far enough from
the black hole will always see a decreasing effect of the plasma on the shadow.
In a homogeneous plasma, $k=0$, however, for any observer in the domain
$3m < r_{\mathrm{O}} < \infty$ the plasma has an increasing effect on the shadow.}
\label{fig:r0k}
\end{figure}

The condition for an increasing effect of the plasma on the shadow can be rewritten 
in this case as
\begin{equation}
x_\mathrm{O}^{k+1} - 3^{k+1} x_\mathrm{O} + 2 \cdot 3^{k+1} < 0 \, , \quad 
x_\mathrm{O} = \frac{r_\mathrm{O}}{m} \, .
\label{eq:r0k}
\end{equation}
Let us restrict ourselves to observers that are farther away from the black hole 
than the photon sphere ($r_\mathrm{O} > 3m$). Then the condition (\ref{eq:r0k}) 
means that for every choice of $k$ there is a limiting value 
$r_\mathrm{O}^{\mathrm{min}}(k)$: if $3m < r_\mathrm{O} < r_\mathrm{O}^{\mathrm{min}}(k)$, 
the shadow becomes bigger due to the presence of the plasma in comparison with the 
vacuum case; if $r_\mathrm{O}^{\mathrm{min}}(k) < r_\mathrm{O}$, it becomes smaller. 
The dependence of $r_\mathrm{O}^{\mathrm{min}}(k)$ on $k$ is shown in Fig. \ref{fig:r0k}.

%----------------------------------------------------------------------------------
\section{The Kerr shadow for some specific plasma distributions}\label{sec:examples}

In this section we discuss a few examples of plasma distributions 
on the Kerr spacetime that satisfy the separability condition (\ref{eq:sepcon}). 
An interesting example would be the case that the plasma frequency depends
on $r$ and $\vartheta$ as the mass density of a dust that is at rest 
at infinity. The analytical form of this matter distribution was found by 
Shapiro \cite{Shapiro1974} whose calculation even included a non-zero pressure.
If one specializes his result to the pressureless case, one finds that the
mass density is independent of $\vartheta $ and proportional to $r^{-3/2}$. In this case the separability condition (\ref{eq:sepcon}) is \emph{not}
satisfied on a Kerr spacetime with $a \neq 0$, which means that our way of calculation can not be directly applied. However, our analytical 
formulas can be applied if we assume an additional $\vartheta$ dependence
of the plasma frequency, see Example 3 below. For the Schwarzschild case,
$a=0$, the shadow in a plasma with plasma frequency proportional to $r^{-3/2}$
was discussed in our earlier paper \cite{PerlickTsupkoBK2015}.

In all the examples we are going to treat in this section we consider a plasma
frequency that satisfies the separability condition (\ref{eq:sepcon}) with
\begin{equation}\label{eq:power}
f_r(r) = C \,  r^k  \, , \quad
f_\vartheta(\vartheta) \ge 0
\end{equation}
where $C \ge 0$ and $0 \le k \le 2$. We 
will now show that in all these cases (\ref{eq:Kspher}) can hold only with the plus sign.
We first treat the case $C=0$. If we assume that in this case (\ref{eq:Kspher}) holds 
with the minus sign we find that $K=0$. By (\ref{eq:ptheta}) this requires 
$f_{\vartheta} ( \vartheta ) =0$, i.e., we are in the vacuum case for which the 
solution set of (\ref{eq:photonreg}) with the lower sign is empty. We now treat the 
case $C>0$. 
We know from (\ref{eq:ptheta}) that $K - f_\vartheta(\vartheta) \ge 0$, therefore 
our assumption $f_{\vartheta} ( \vartheta ) \ge 0$ implies that $K$ is non-negative.
Let us assume that (\ref{eq:Kspher}) holds with the minus sign and write $K$ as
\begin{equation}\label{eq:Kx}
K= \frac{r^2 \Delta \omega_0^2}{(r-m)^2} \left( 1-\sqrt{1-x} \right)^2 - f_r(r)
\end{equation}
with
\begin{equation}\label{eq:x}
x = f'_r(r) \frac{r-m}{2r^2 \omega_0^2}, \quad \text{and}  \quad 0 \le x \le 1\, .
\end{equation}
It can be easily shown that
\begin{equation}\label{eq:xest}
\left( 1-\sqrt{1-x} \right)^2 \le x^2 \le x \quad \text{for} \quad 0 \le x \le 1 \, .
\end{equation}
Hence
\begin{gather}\label{eq:Kest}
K= \frac{r^2 \Delta \omega_0^2}{(r-m)^2} \left( 1-\sqrt{1-x} \right)^2 - f_r(r) 
\\
\nonumber
\le \, 
\frac{r^2 \Delta \omega_0^2}{(r-m)^2} \, x \, -  \, f_r(r) 
= 
\frac{1}{2} \frac{\Delta f'_r(r)}{r-m} \, -  \, f_r(r) \, .
\end{gather}
We see that $\omega_0^2$ has dropped out. As, by (\ref{eq:power}),
\begin{equation}\label{eq:fpest}
f'_r(r) = k \, \frac{f_r(r)}{r} \, ,
\end{equation}
this implies
\begin{equation}\label{eq:Kk}
K  \, \le  \,  \frac{k}{2}  \, \frac{\Delta f_r(r)}{r(r-m)} - f_r(r) \, .
\end{equation}
As we assume $0 \le k \le 2$, this implies
\begin{gather}\label{eq: Kk2}
K \, \le \, \frac{\Delta f_r(r)}{r(r-m)} - f_r(r)  
\\
\nonumber
= \, \frac{(r^2 - 2mr +a^2) f_r(r)}{r(r-m)} - f_r(r) 
\\
\nonumber
\le \, 
\frac{(r^2 - 2mr +m^2) f_r(r)}{r(r-m)} - f_r(r) 
\\
\nonumber
= \, \frac{r-m}{r}  \, f_r(r)  - f_r(r) \, = \, - \, \frac{m}{r} \,  f_r(r)  \, <  \,  0 \, .
\end{gather}
We have thus obtained that $K<0$, which is the 
desired contradiction.

Therefore we have to consider (\ref{eq:Kspher}) only with the plus sign in the
following examples.

As explained at the end of Sec.~\ref{sec:shadow}, we use in all 
examples the same method of plotting the shadow as 
in \cite{GrenzebachPerlickLaemmerzahl2014}. When we are
interested in the extreme case, $a=m$, we actually choose $a=0.999 m$ for our 
plots because our analytical formulas for the boundary curve of the shadow
involve undetermined expressions when 
$a$ is exactly equal to $m$. Of course, the pictures of the shadow for $a=0.999m$ are
practically indistinguishable from those for $a=m$.

\vspace{0.6cm}

%--------------------------------------------------------------------------------

\noindent
\emph{Example $1$: The vacuum shadow on Kerr spacetime}

\begin{figure}[h]
\includegraphics[width=0.45\textwidth]{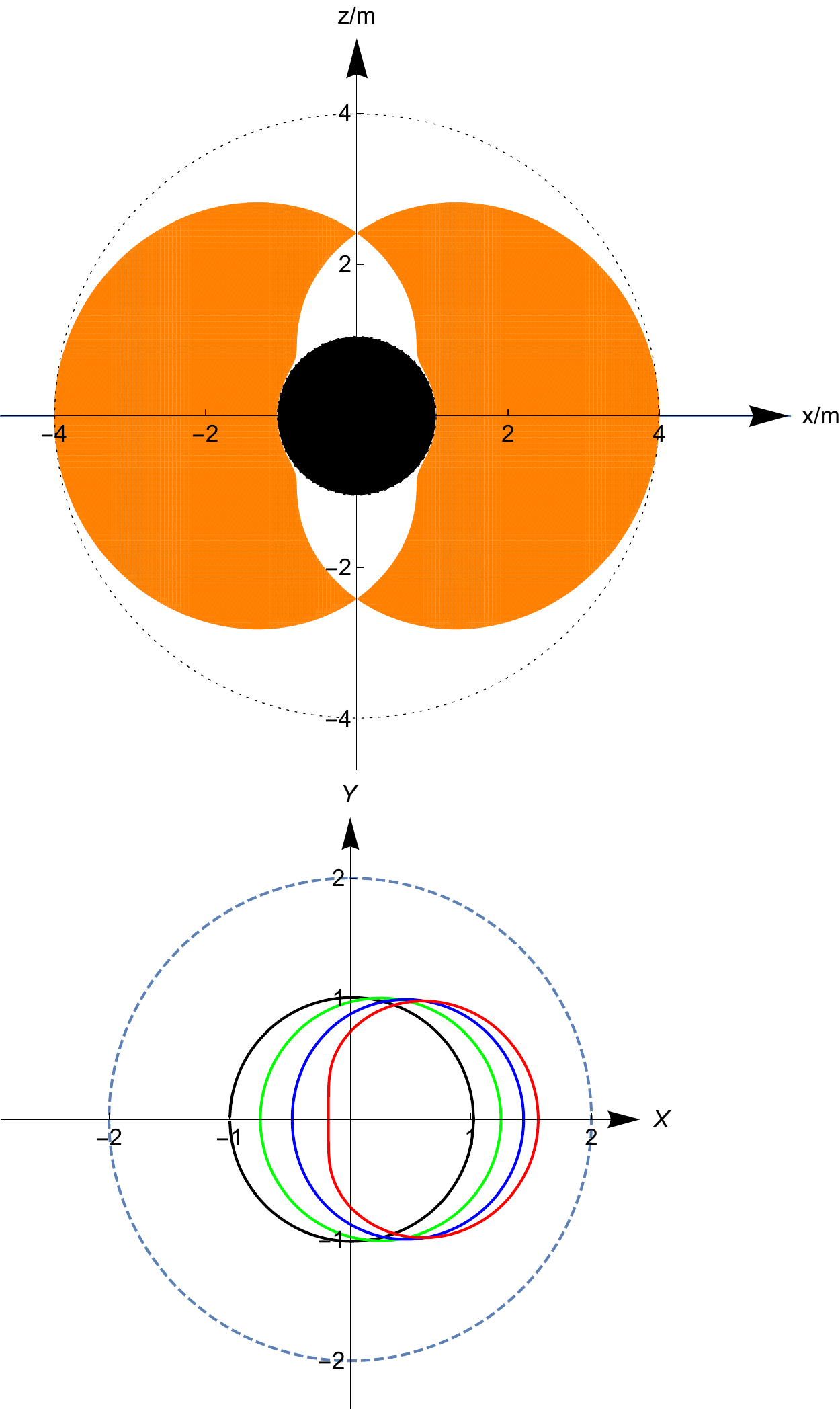}
\caption{Example 1, $\omega _p =0$ (vacuum). 
Top: The photon region is shown for a black hole with $a=0.999 \, m$. 
Here and in all the following pictures, the black disk is the region inside the 
horizon. For the chosen value of $a$ it is almost identical with the sphere
of radius $m$. We use coordinates $x$, $y$ (not shown) and $z$ related to the
Boyer-Lindquist coordinates by the usual transformation formulas from spherical
polars to Cartesian coordinates. The dotted circles (here and in all the 
following pictures) intersect the boundary of the photon region where 
circular light rays in the equatorial plane exist; the one on the inner boundary 
is co-rotating while the one on the outer boundary is counter-rotating. For the 
chosen value of $a$ in vacuum the inner one is at a radius only slightly bigger
than the radius of the horizon. Bottom:
The picture shows the shadow for an observer at $r_{\mathrm{O}}= 5 \, m$ and 
$\vartheta _{\mathrm{O}} = \pi /2$, with the spin parameter 
$a$ equal to $0.02 \, m$ (leftmost, black), $0.4 \, m$ (green), $0.75 \, m$ (blue) 
and $0.999 \, m$ (red). The axes label $X$ and $Y$ refer to the Cartesian
coordinates (\ref{eq:stereo}). The dashed circle is the celestial equator.
\label{fig:e1}}
\end{figure}

For the sake of comparison, we briefly revisit the known case of the Kerr 
shadow in vacuum, $\omega _p =0$, which is equivalent to the limiting case
of infinite photon frequency, $\omega _0 \to \infty$, in a plasma of 
arbitrary density.
The vacuum case is described by the first terms in formulas (\ref{small-theta}) and 
(\ref{small-psi}). These analytical formulas for the boundary curve of the shadow 
of a Kerr black hole in vacuum can be found as special cases 
in \cite{GrenzebachPerlickLaemmerzahl2014} and \cite{GrenzebachPerlickLaemmerzahl2015}. 
In Fig.~\ref{fig:e1} we plot the photon region of an almost extreme Kerr black hole 
in vacuum and the shadow for different parameters $a$.

\vspace{0.6cm}

%--------------------------------------------------------------------------------
\noindent
\emph{Example $2$: A plasma density with $f_r(r) =0$ on Kerr spacetime}

As an example, we consider a plasma frequency which satisfies the 
separability condition with $f_r'(r) =0$. As a constant $f_r(r)$ may be 
incorporated into $f_{\vartheta} ( \vartheta )$, we may choose $f_r (r) =0$,
i.e.
\begin{equation}\label{eq:fr=0}
\omega _p (r , \vartheta ) ^2 = 
\frac{f_{\vartheta} ( \vartheta )}{r^2 + a^2 \mathrm{sin} ^ 2 \vartheta }
\, .
\end{equation}

Then we must have $f_{\vartheta } ( \vartheta ) \ge 0$ and, as shown in the
introductory part of this section, the 
inequality (\ref{eq:photonreg}) can hold only with the upper sign, 
i.e., the photon region is given by
\begin{equation}\label{eq:photonregfr=0}
\left( \frac{4 \, r^2 \Delta}{(r-m)^2} - 
\frac{f _{\vartheta} ( \vartheta )}{\omega _0^2} \right) 
 a^2 \mathrm{sin} ^2 \vartheta \ge 
\end{equation} 
\[
\ge
\left( \frac{2}{(r-m)} \Big(m(a^2-r^2)+ r \, \Delta \Big) - a^2 \mathrm{sin} ^2 \vartheta \right) ^2
\, .
\]
By (\ref{eq:pphispher}) and (\ref{eq:Kspher}), respectively, the constants of motion
of a spherical light ray at $r = r_p$ are given by
\begin{equation}\label{eq:pphifr=0} 
a p_{\varphi}  = - \, \dfrac{\omega _0}{(r-m)} 
\Big( m(a^2-r^2) + 
r \, \Delta  \Big)
\, ,
\end{equation}
\begin{equation}\label{eq:Kfr=0} 
K  =  \dfrac{4 \, r^2 \Delta \omega _0 ^2}{(r-m)^2} 
\, .
\end{equation}

For the pictures in Fig.~\ref{fig:e2} we have chosen
$f_{\vartheta} ( \vartheta ) = 
\omega _c ^2 m^2 (1+ 2 \, \mathrm{sin} ^2 \vartheta )$, i. e.
\begin{equation}\label{eq:ex2}
\omega _p (r , \vartheta )^2 \, = \, 
\frac{
\omega _c ^2 m^2 (1+ 2 \, \mathrm{sin} ^2 \vartheta )
}{
r^2 +a^2 \mathrm{cos}^2 \vartheta
}
\end{equation}
where $\omega _c$ is a constant with the dimension of a frequency.

As in the vacuum case, all spherical light rays in the domain of 
outer communication are unstable, so they may serve as limit
curves for light rays issuing from the observer position into the
past. Recall that we have to consider the intersection of the 
unstable photon region with the cone $\vartheta _{\mathrm{O}}
= \mathrm{constant}$ for constructing the boundary curve of the
shadow, where $\vartheta _{\mathrm{O}}$ is the $\vartheta$
coordinate of the observer. In Fig. \ref{fig:e2} we see that 
the photon region becomes detached from the equatorial plane
at a certain critical value of $\omega _c^2 / \omega _0^2$ and
that a region forms where light rays with frequency value
$\omega _0$ are forbidden. If this happens, observers with 
$\vartheta _{\mathrm{O}}$ near $\pi /2$ do not see a shadow 
anymore. If they are farther away from the black hole as the
forbidden region, they see a completely bright sky; if they
are between the forbidden region and the black hole, their
sky is dark.

\vspace{0.6cm}

%--------------------------------------------------------------------------------
\noindent
\emph{Example $3$: A plasma density with $f_r(r) \propto \sqrt{r}$ 
            and $f_{\vartheta} ( \vartheta ) =0$ on Kerr spacetime}

As our next example, we consider an inhomogeneous plasma where the plasma 
density is proportional to $r^{-3/2}$, as for a dust; however, in order
to satisfy the separability condition we have to assume a $\vartheta$ dependence in
addition,
\begin{equation}\label{eq:ex4}
\omega _p ( r , \vartheta )^2 = 
\frac{\omega _c ^2 \, \sqrt{m^3r}}{r^2 + a^2 \mathrm{cos} ^2 \vartheta}
\end{equation}
with a constant $\omega _c$ that has the dimension of a frequency. As in Example 2,
all spherical light rays are unstable and a forbidden region is formed when
$\omega _c^2 / \omega _0^2$ becomes bigger than a certain critical 
value, see Figure \ref{fig:e3}. However, in contrast to Example 2 now the
photon region becomes detached from the axis, rather than from the 
equatorial plane. As a consequence, for certain values of $\omega _0$ observers
near the equatorial plane still see a shadow while observers near the 
axis do not. If $\omega _c^2 / \omega _0 ^2$ becomes too big, the shadow
vanishes for all observers.

\vspace{0.6cm}

%--------------------------------------------------------------------------------
\noindent
\emph{Example $4$: A homogeneous plasma on Kerr spacetime}

Finally let us consider the case of a homogeneous plasma, $\omega _p ( r , \vartheta )
= \omega _c = \mathrm{constant}$, for which the functions $f_r (r)$ and 
$f_{\vartheta} ( \vartheta )$ are given by (\ref{eq:hom}). The new feature, in 
comparison to the previous examples, is the existence of \emph{stable}
spherical light rays. This implies that from some observer
positions there are light rays issuing into the past that go neither
to infinity nor to the horizon; 
they rather stay inside a spatially compact region. If we 
stick with the rule that we assign darkness only to those past-oriented 
light rays that go to the horizon, we have to assign brightness
to these light rays. The existence of bound photon orbits in a 
homogeneous plasma was also discussed by Kulsrud and Loeb \cite{KulsrudLoeb1992}
and by Bisnovatyi-Kogan and Tsupko \cite{BKTs2010}. Note that stable 
circular light rays exist also in some non-homogenous plasma distributions,
see Rogers \cite{Rogers2017b}.

Figure \ref{fig:e4} shows the development of the stable and unstable 
photon regions and of the forbidden region. In this case the presence of 
the plasma always makes the shadow larger, in comparison to an observer at the 
same coordinate position in vacuum, see Fig.~\ref{fig:e4sh}. 
This is similar to the Schwarzschild case, see 
Sec.~\ref{sec:Schwarzschild}. If $\omega _c^2 / \omega _0 ^2$
reaches a certain limit value, the shadow covers the entire sky.

\vspace{0.6cm}

%-----------------------------------------------------------------------
\section{Conclusions}

In this work we have investigated the propagation of light rays in a 
non-magnetized pressureless plasma on Kerr spacetime. We have worked in the
frame-work of geometrical optics, considering the plasma as a medium 
with dispersive properties given by a frequency-dependent index of 
refraction. The gravitational field is determined by the mass and the
spin of the Kerr black hole, i.e., the gravitational field of the 
plasma particles is not taken into account. In our model, the presence of 
the plasma manifests itself only in a change of the trajectories of light 
rays. Our investigation is valid for any value of the spin parameter $a$ 
and for any light rays in the domain of outer communication. As the plasma 
is a dispersive medium, photons with different frequencies move along different 
trajectories. Therefore, all phenomena described here are chromatic.

We have utilized the Hamiltonian formalism for light rays in a plasma on 
Kerr spacetime and we have, in an important first step of our analysis, 
determined the necessary and sufficient condition for separability of
the Hamilton-Jacobi equation, see Secs.~\ref{sec:Hamilton} and \ref{sec:sepcon}. 
We have demonstrated that the Hamilton-Jacobi equation is separable, i.e., that a
generalized Carter constant exists, only for special distributions of the 
plasma electron density. The necessary and sufficient condition for 
separability is given in eq.~(\ref{eq:sepcon}).

For all cases where the condition of separability is satisfied we have determined 
the photon region, i.e., the region in spacetime where spherical light rays exist, 
see Sec.~\ref{sec:phreg}. Examples of photon regions are given in 
Sec.~\ref{sec:examples}.

We have derived analytical formulas for the boundary curve of the shadow on 
the observer’s sky in terms of two angular celestial coordinates, see formulas 
(\ref{eq:shtheta}) and (\ref{eq:shpsi}) with (\ref{eq:pphispher}) and (\ref{eq:Kspher}). 
Our formulas are valid for any photon frequency at infinity $\omega _0$, for any value 
of the spin parameter $a$, for any position of the observer outside of the horizon 
of the black hole (i.e., with arbitrary distance from the black hole and arbitrary 
inclination) and for any plasma distribution which satisfies the separability 
condition. For the reader's convenience, we have written a step-by-step procedure 
for the construction of the shadow, see the end of Sec.~\ref{sec:shadow}.
We have worked out several examples, considering a low-density plasma in 
Sec.~\ref{sec:approx}, a non-rotating black hole in Sec.~\ref{sec:Schwarzschild}
and several specific plasma distributions on the Kerr spacetime in Sec.~\ref{sec:examples}.

In our approach, the plasma influences only the trajectories of light rays which
results in a change of the geometrical size and shape of the shadow. We have not 
taken into account processes of absorption or scattering of photons; there are 
several ray tracing codes for light propagation in matter on the Kerr spacetime
that take such effects into account, see e.g. 
James et al.~\cite{JamesTunzelmannFranklinThorne2015}. 
In this paper, however, it was our goal to derive an analytical formula for the 
boundary curve of the shadow under idealized conditions. Our equations may be used as 
the zeroth order approximation for numerically studying more realistic situations.

In the presence of a plasma around a black hole, the size and the shape of the 
shadow differs from the vacuum case in  a way that depends on the photon frequency. 
The relevant quantity is the ratio of the plasma frequency to the photon frequency.
From our examples we conclude that if the plasma frequency is small in
comparison with the photon frequency, the shadow is not very much different from
the vacuum case. However, if the plasma frequency is close to the photon frequency the
properties of the shadow are changed drastically because of significant changes
in the photon regions.
 
In all applications to astrophysics, the difference of the plasma case to the
vacuum case is significant only for radio frequencies. For an estimate of the
effects  in the case of Sgr A$^*$ and M87 we refer to our previous 
paper~\cite{PerlickTsupkoBK2015}. At low frequencies, 
i.e. at large wavelengths, where the influence of 
the plasma is most significant, scattering is expected to be non-negligible.
As scattering will partly wash out the shadow, it is expected that the shadow
can be observed only at wavelengths of approximately one millimeter or below
where the plasma effects are very small.

%-----------------------------------------------------------------------
\section*{Acknowledgments}

It is our pleasure to thank Gennady S. Bisnovatyi-Kogan for
helpful discussions and permanent interest in our research. VP gratefully acknowledges
support from the Deutsche Forschungsgemeinschaft
within the Research Training Group 1620 ``Models of
Gravity.'' The results for the shadow with low density and power-law models of plasma were obtained by O. Yu. T., and this part of work was financially supported by Russian Science Foundation, Grant No. 15-12-30016.

\bibliographystyle{ieeetr}

\begin{widetext}
\begin{center}
\begin{figure}[h]
\includegraphics[width=0.8\textwidth]{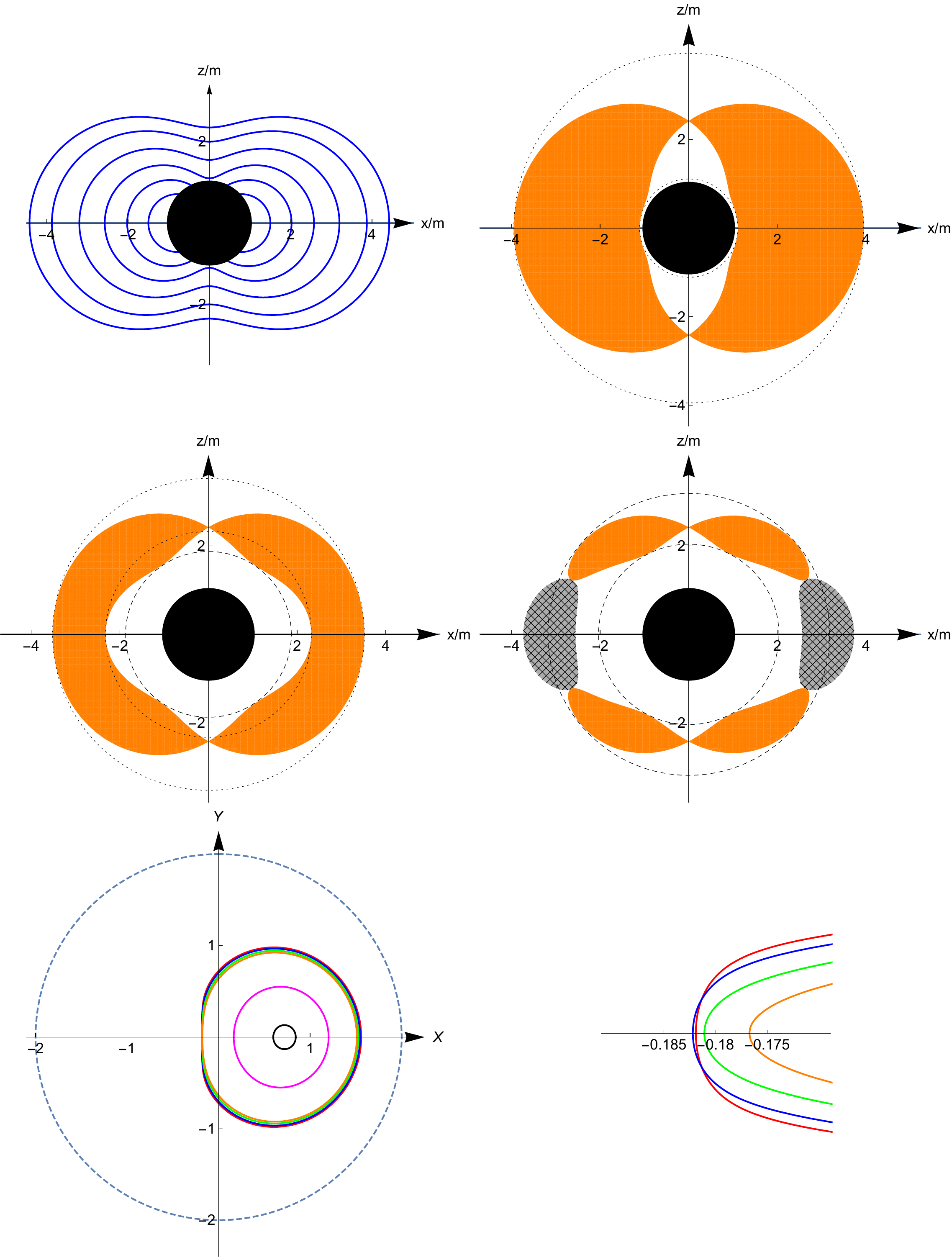}
\caption{Example 2, $\omega _p (r , \vartheta )^2 =
\omega _c ^2 m^2 (1+ 2 \, \mathrm{sin} ^2 \vartheta )
/
(r^2 +a^2 \mathrm{cos}^2 \vartheta)$. Top left: The surfaces $\omega _p =
\mathrm{constant}$. The next three pictures show the photon region for 
$a = 0.999 \, m$. Top right: $\omega _c^2/\omega _0^2 = 1$. The situation
is qualitatively similar to the vacuum case, cf. Fig. \ref{fig:e1}.
Middle left: $\omega _c^2/\omega _0^2 = 7$. In addition to the two 
circular light rays in the equatorial plane (located where the dotted
circles meet the photon region) there is a pair of circular light rays 
off the equatorial plane (located where the dashed circle meets the
photon region). Middle right: $\omega _c^2/\omega _0^2 = 10$. At
$\omega _c^2/\omega _0^2 \approx 9.000$ the photon region  has become
detached from the equatorial plane. A region (cross-hatched) has 
formed which is forbidden for light rays with the chosen $\omega _0$.
Observers close to the equatorial plane do no longer see a shadow.
At $\omega _c^2 / \omega _0^2 \approx 23.314$ the photon region
vanishes, the forbidden region encloses the black hole completely. 
Bottom left: Boundary curve of the shadow for an observer at 
$r_{\mathrm{O}} = 5 \, m$ and 
$\vartheta _{\mathrm{O}} = \pi /2$ with $a = 0.999 \, m$ and 
$\omega _c^2 /\omega _0^2$ equal to 0 (outermost, red), 0.4 (blue),
1 (green), 1.5 (orange), 7 (magenta) and 8.9 (black). The shadow has shrunk to
a point when the forbidden region reaches the observer position.
Bottom right: Enlarged view of the first four boundary curves,
stretched in the horizontal direction by a factor of 40.\label{fig:e2}}
\end{figure}

\begin{figure}[h]
\includegraphics[width=0.8\textwidth]{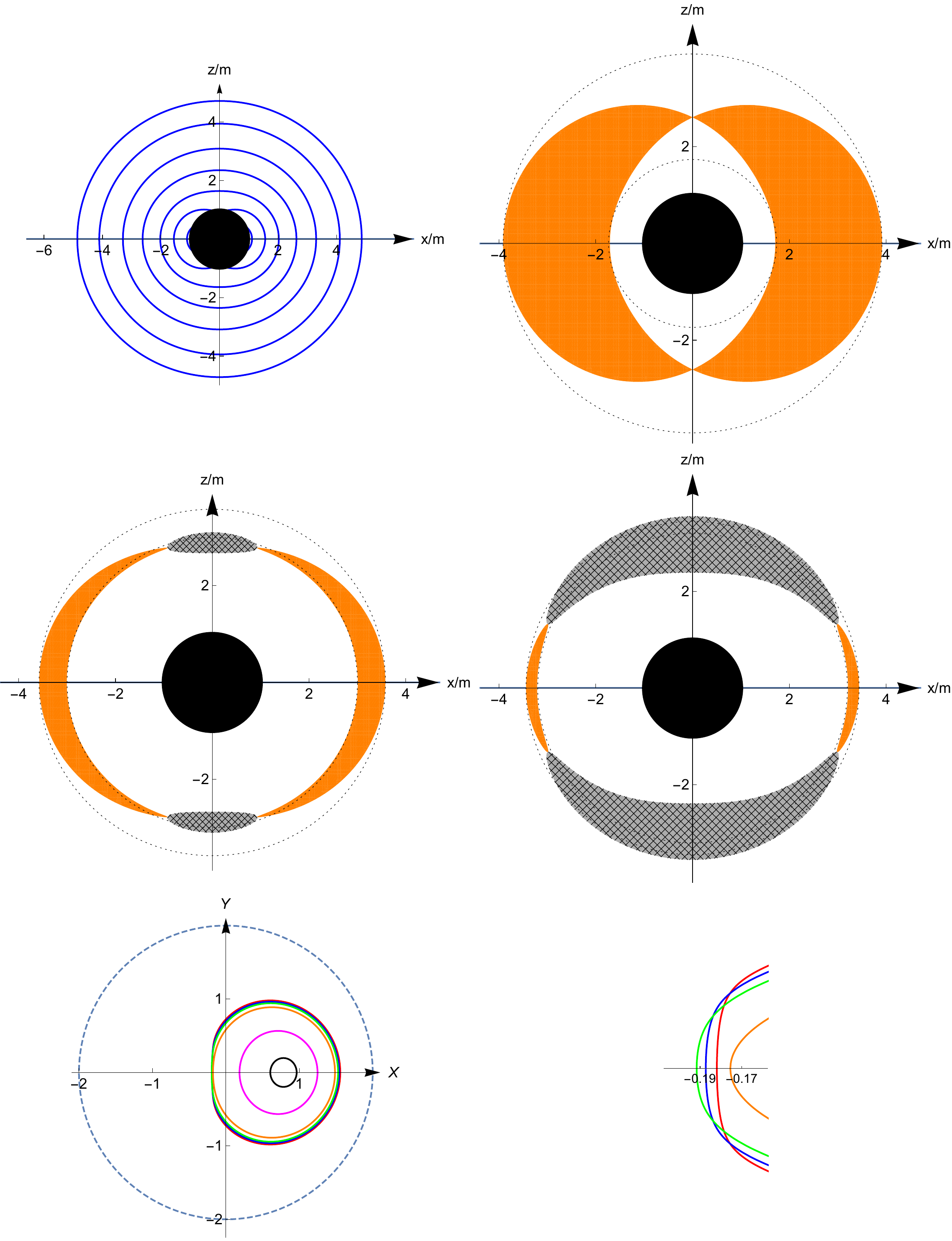}
\caption{Example 3, $\omega _p (r , \vartheta )^2 =
\omega _c ^2 \, \sqrt{m^3r}/(r^2 + a^2 \mathrm{cos} ^2 \vartheta )$.
Top left: The surfaces $\omega _p =
\mathrm{constant}$. The next three pictures show the photon region for 
$a = 0.999 \, m$. Top right: $\omega _c^2/\omega _0^2 = 7$. The situation
is qualitatively similar to the vacuum case, cf. Fig. \ref{fig:e1}.
Middle left: $\omega _c^2/\omega _0^2 = 14.5$. At $\omega _c^2/\omega _0^2 
\approx = 14.402$ the photon region has become detached from the axis.
A region (cross-hatched) has formed which is forbidden for light rays 
with the chosen $\omega _0$. Observers close to the axis do no longer 
see a shadow. Middle right: $\omega _c^2/\omega _0^2 = 15.1$. The
photon region has further shrunk. At $\omega _c^2/\omega _0^2 \approx 15.215$ 
it has vanished and the forbidden region encloses the black hole completely. 
Bottom left: Boundary curve of the shadow for an observer at 
$r_{\mathrm{O}} = 5 \, m$ and 
$\vartheta _{\mathrm{O}} = \pi /2$ with $a = 0.999 \, m$ and 
$\omega _c^2 /\omega _0^2$ equal to 0 (outermost, red), 1 (blue),
3 (green), 6 (orange), 13 (magenta) and 15 (black). The shadow has shrunk to
a point when the forbidden region reaches the equatorial plane.
Bottom right: Enlarged view of the first four boundary curves,
stretched in the horizontal direction by a factor of 40.\label{fig:e3}}
\end{figure}

\begin{figure}[h]
\includegraphics[width=0.8\textwidth]{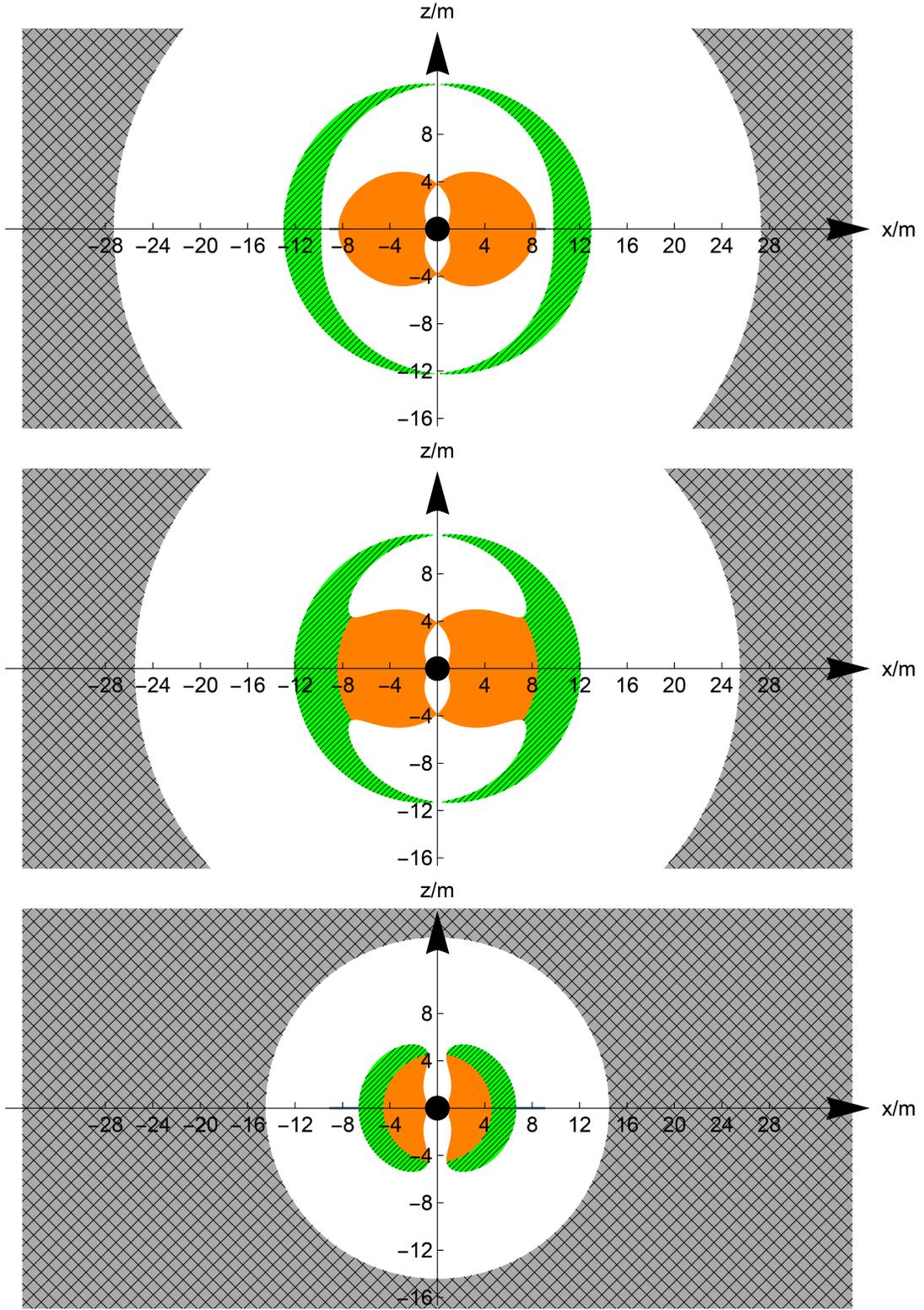}
\caption{Example 4, $\omega _p (r , \vartheta )^2=
\omega _c ^2$. The pictures show the photon region  for 
$a = 0.999 \, m$ and $\omega _c^2 / \omega _0 ^2$ equal to
1.079 (top), 1.085 (middle) and 1.16 (bottom). For $\omega _c ^2 /
\omega _0 ^2 < 1$ the photon region is qualitatively similar to the vacuum 
case. At $\omega _c^2 / \omega _0^2 =1$ a forbidden region (cross-hatched)
and a region with stable spherical light rays (hatched, green) come
into existence. At $\omega _c^2 / \omega _0^2 \approx 1.080$ the stable
and the unstable photon regions come together. At $\omega _c^2 / \omega _0^2 
\approx 1.14$ the photon region becomes detached from the axis. At
$\omega _c^2 / \omega _0^2 \approx 2.29$ the photon region vanishes.
\label{fig:e4}}
\end{figure}

\begin{figure}[h]
\includegraphics[width=0.8\textwidth]{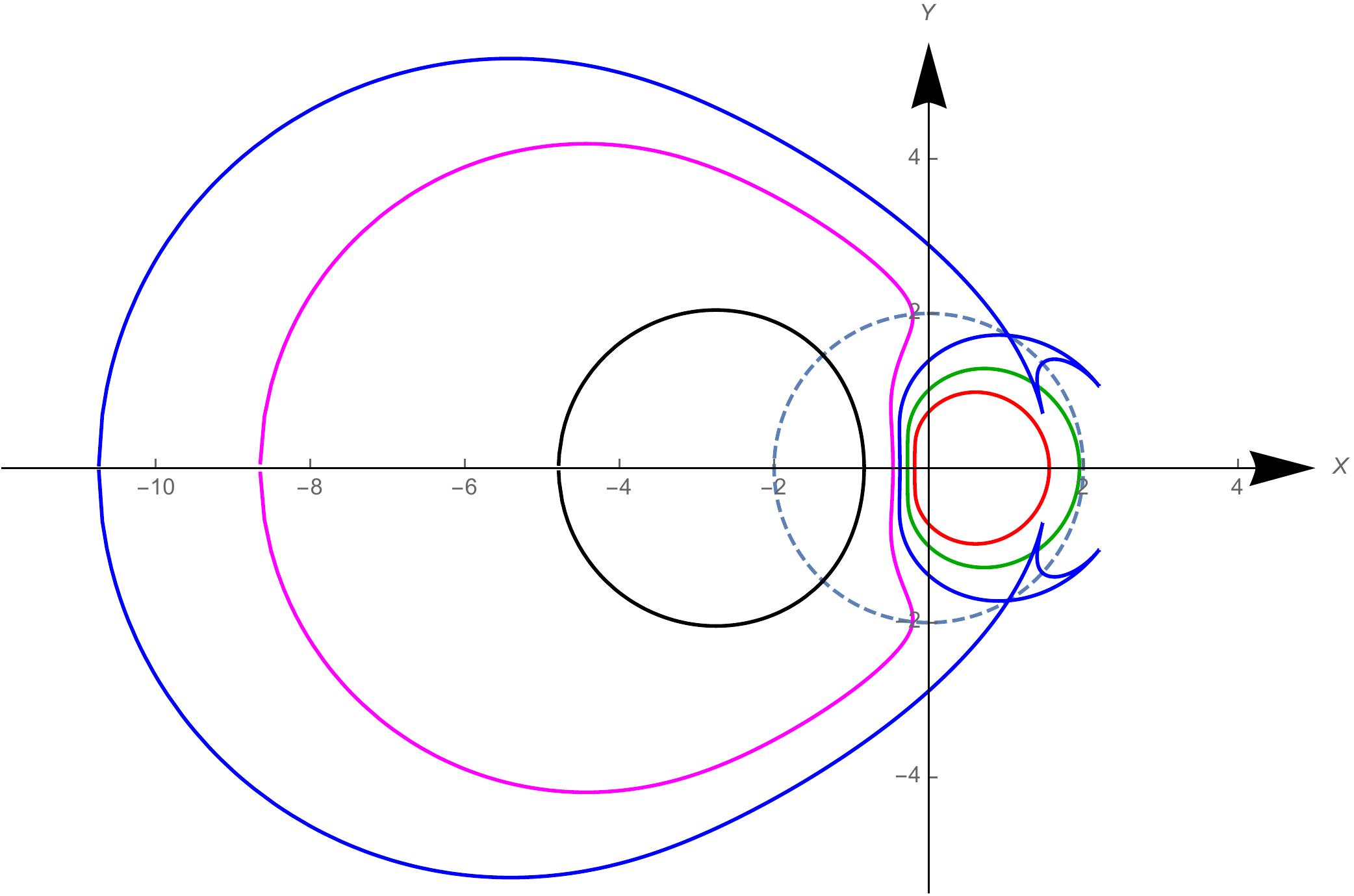}
\caption{Example 4, $\omega _p (r , \vartheta )^2=
\omega _c ^2$. The picture shows the boundary curve of
the shadow for an observer at $r_{\mathrm{O}} = 5 \, m$ and 
$\vartheta _{\mathrm{O}} = \pi /2$ with $a= 0.999 \, m$ and
$\omega _c^2 /\omega _0^2$ equal to 0 (innermost, red), 
0.8 (green), 1.085 (blue), 1.2 (magenta) and 1.345 (black). 
When the forbidden region is formed at $\omega _c^2 / \omega _0^2 =1$,
the shadow includes the point $\theta = \pi$ (i.e., the 
direction pointing away from the black hole that corresponds
to the point at infinity in the stereographic projection). 
Then, for a certain range of values
of $\omega _c^2 / \omega _0^2$, the shadow displays two 
structures similar to fish-tails, bright inside and dark outside. 
If $\omega _c^2 / \omega _0^2$
is further increased, the shadow grows until it covers the 
entire sky.
\label{fig:e4sh}}
\end{figure}

\end{center}
\end{widetext}

\end{document}